\documentclass[lettersize,journal]{IEEEtran}
\usepackage{amsmath,amsfonts}
\usepackage{algorithmic}
\usepackage{algorithm}
\usepackage{array}
\usepackage[caption=false,font=footnotesize,labelfont=rm,textfont=rm]{subfig}
\usepackage{textcomp}
\usepackage{stfloats}
\usepackage{url}
\usepackage{verbatim}
\usepackage{graphicx}
\usepackage{cite}
\usepackage{color}

\usepackage{amsmath} 
\RequirePackage{bibspacing}

\AtBeginDocument{
    \setlength{\jot}{1.2pt}
    \setlength{\abovedisplayskip}{1.2pt}
    \setlength{\abovedisplayshortskip}{1.2pt}
    \setlength{\belowdisplayskip}{1.2pt}
    \setlength{\belowdisplayshortskip}{1.2pt}
}

\usepackage[linkcolor=red,
            anchorcolor=blue,
            citecolor=green]{hyperref}

\usepackage{cleveref}

\usepackage{xcolor}
\usepackage{xpatch}

\makeatletter
\ExplSyntaxOn
\cs_new:Npn \bibColoredItems #1#2
  {
    \clist_map_inline:nn {#2} { \cs_new:cpn {bib@colored@##1} {#1} } 
  }
\ExplSyntaxOff

\newcommand\bib@setcolor[1]{%
  \ifcsname bib@colored@#1\endcsname
    \expanded{\noexpand\color{\csname bib@colored@#1\endcsname}}%
  \else
    \normalcolor
  \fi
}

\IfPackageLoadedTF{hyperref}{\@tempswatrue}{\@tempswafalse}
\if@tempswa
  \xpatchcmd\@bibitem {\H@item}{\bib@setcolor{#1}\H@item}{}{\PatchFailed}
  \xpatchcmd\@lbibitem{\H@item}{\bib@setcolor{#2}\H@item}{}{\PatchFailed}
\else
  \xpatchcmd\@bibitem {\item}  {\bib@setcolor{#1}\item}  {}{\PatchFailed}
  \xpatchcmd\@lbibitem{\item}  {\bib@setcolor{#2}\item}  {}{\PatchFailed}
\fi
\makeatother

\definecolor{revisionblue}{HTML}{0000FF}

\begin{document}

\title{Dependency-Aware Task Offloading in Multi-UAV Assisted Collaborative Mobile Edge Computing}

\author{Zhenyu Zhao,~\IEEEmembership{Graduate Student Member, IEEE}, Xiaoxia Xu, Tiankui Zhang,~\IEEEmembership{Senior Member, IEEE}, \par
Junjie Li, Yuanwei Liu,~\IEEEmembership{Fellow, IEEE}
\thanks{Zhenyu Zhao and Tiankui Zhang are with the School of Information and Communication Engineering, Beijing University of Posts and Telecommunications, Beijing 100876, China (e-mail: zhaozhenyu@bupt.edu.cn; zhangtiankui@bupt.edu.cn).}

\thanks{Xiaoxia Xu is with the School of Electronic Information, Wuhan University, Wuhan 430072, China (e-mail: xiaoxiaxu@whu.edu.cn).}

\thanks{Junjie Li is with China Telecom Beijing Research Institute, Beijing 102200, China (e-mail: lijj28@chinatelecom.cn).}

\thanks{Yuanwei Liu is with the School of Electronic Engineering and Computer Science, Queen Mary University of London, E1 4NS London, U.K. (e-mail: yuanwei.liu@qmul.ac.uk).}

}

\markboth{Journal of \LaTeX\ Class Files,~Vol.~14, No.~8, August~2021}%
{Shell \MakeLowercase{\textit{et al.}}: Dependency-Aware Task Offloading in Multi-UAV Assisted Collaborative Mobile Edge Computing}


\maketitle

\begin{abstract}

This paper proposes a novel multi-unmanned aerial vehicle (UAV) assisted collaborative mobile edge computing (MEC) framework, where the computing tasks of terminal devices (TDs) can be decomposed into serial or parallel sub-tasks and offloaded to collaborative UAVs. We first model the dependencies among all sub-tasks as a directed acyclic graph (DAG) and design a two-timescale frame structure to decouple the sub-task interdependencies for sub-task scheduling. Then, a joint sub-task offloading, computational resource allocation, and UAV trajectories optimization problem is formulated, which aims to minimize the system cost, i.e., the weighted sum of the task completion delay and the system energy consumption. To solve this non-convex mixed-integer nonlinear programming (MINLP) problem, a penalty dual decomposition and successive convex approximation (PDD-SCA) algorithm is developed. Particularly, the original MINLP problem is equivalently transferred into a continuous form relying on PDD theory. By decoupling the resulting problem into three nested subproblems, the SCA method is further combined to recast the non-convex components and obtain desirable solutions. Numerical results demonstrate that: 1) Compared to the benchmark algorithms, the proposed scheme can significantly reduce the system cost, and thus realize an improved trade-off between task latency and energy consumption; 2) The proposed algorithm can achieve an efficient workload balancing for distributed computation across multiple UAVs.

\end{abstract}


\section{Introduction}
\IEEEPARstart{T}{he} popularization of terminal devices (TDs) and the emergence of various mobile applications place high demands on the communication and computation capabilities of future wireless networks. In order to alleviate the heavy burden on the network caused by the communication and computation demands of large-scale TDs, Mobile Edge Computing (MEC) technology has been widely used~\cite{ref2}. By providing computing services at the edge of the mobile network, MEC enables localized and nearby deployment of computing services, thus reducing task completion delay and alleviating the pressure on the original network.\par

In large-scale scenarios, it is difficult to rely only on a single edge server to take up the computational demands of multiple tasks~\cite{CollMEC1}. To overcome this problem, collaborative MEC is introduced, in which multiple edge servers work together and share resources to efficiently accomplish computationally intensive tasks. Collaboration of multiple edge servers not only improves the MEC service capability, but also reduces the task completion latency~\cite{CollMEC2}.\par

To benefit from collaborative MEC, it is crucial to investigate effective task offloading and resource allocation (communication and computational resource) methods. For collaborative MEC scenarios, adjusting task offloading and resource allocation based on task characteristics and network conditions can optimize task completion latency, system energy consumption, and server computational load balance. Extensive research has been conducted on the joint optimization of computational and communication resource allocation and task offloading in collaborative MEC scenarios~\cite{CollMEC2,ref3,ref4,ref5}. The task offloading methods in these studies include binary offloading methods~\cite{ref3,ref4} as well as partial offloading methods~\cite{ref4,ref5}. The binary offloading method offloads a single task to a single MEC server for computation, while the partial offloading method subdivides the task into smaller sub-tasks and offloads them to multiple MEC servers for computation. Compared to binary offloading methods, partial offloading methods can more efficiently utilize system resources, balance server workloads, and reduce task completion latency~\cite{ref4}. In addition, due to the emergence of virtualization technologies (e.g., containers and virtual machines) and the use of microservices architectures in MECs, decomposing large-scale computation tasks into lightweight sub-tasks has become a popular and practical solution~\cite{ref8-1}.\par

\subsection{Related Work}
\subsubsection{Task Offloading for Terrestrial MEC}

Previous partial offloading approaches divide tasks into independent sub-tasks, which improves system resource utilization and reduces task completion latency, but ignores the dependencies between sub-tasks~\cite{ref4,ref5}. For example, in computation-intensive tasks such as Artificial Intelligence (AI) image processing, computational steps are usually interrelated, and the output result of one step needs to be passed as an input to the next step.\par

In order to model the dependencies between sub-tasks, existing research has introduced the Directed Acyclic Graph (DAG), which models a task as a series of interdependent sub-task chains based on the underlying correlations between sub-tasks~\cite{ref14,ref15,ref11,newDependenyTask1,newDependenyTask2 ,newDependenyTask3,newDependenyTask5}. These research focuses on designing queue networks and sub-task offloading schemes to maximize system throughput~\cite{ref14,ref15}, or optimizing sub-task offloading and resource allocation to reduce system cost (e.g., task completion delay and energy consumption)~\cite{ref11,newDependenyTask1, newDependenyTask2,newDependenyTask3,newDependenyTask5}. Specifically, to maximize system throughput, the study in~\cite{ref14} introduced the virtual queue network to represent the completion states of interdependent sub-tasks, then a sub-task scheduling strategy is designed based on the virtual queue network to maximize the system throughput. Similarly,~\cite{ref15} designed a virtual queue network and corresponding sub-task scheduling scheme based on the analyzed network capacity region to maximize the system throughput. In addition, in the study of optimizing system cost, the authors in~\cite{ref11} modeled the tasks as serial-dependency sub-task chains. Then, a sub-task scheduling algorithm is developed to minimize the task completion delay. Similarly,~\cite{newDependenyTask1} investigated the offloading of mutually independent sub-tasks and serial-dependency sub-tasks, respectively, and designed sub-task scheduling algorithms to minimize the task completion delay and system energy consumption. In addition to decomposing tasks into serial-dependency sub-task chains, some studies consider that parallel relationships can exist between sub-tasks~\cite{newDependenyTask2,newDependenyTask3,newDependenyTask5}. Further subdividing a single sub-tasks into multiple smaller parallel sub-tasks and offloading them to multiple servers allows for better utilization of idle system resources. The research in~\cite{newDependenyTask2} considered the existence of serial dependencies or parallel relationships between sub-tasks, a global sub-task priority calculation method is designed to determine the computation priority of sub-tasks, and a semi-distributed algorithm is proposed to obtain the sub-task offloading scheme. Similarly,~\cite{newDependenyTask3} and~\cite{newDependenyTask5} designed a sub-task scheduling scheme based on sub-task priority calculation. Specifically, the work in~\cite{newDependenyTask3} introduced a topological sorting algorithm to obtain sub-task scheduling priorities. The work in~\cite{newDependenyTask5} introduced the Reverse Breadth First Search (RBFS) algorithm to generate the priority of each sub-task.\par

\subsubsection{Task Offloading for UAV-assisted MEC}

All of the above studies on dependency-aware sub-task offloading in MEC have focused on ground-based MEC scenarios, whereas ground-based MEC servers have limited accessibility in specific scenarios (e.g., disaster areas and remote wilderness environments). In contrast, UAVs can be deployed flexibly to quickly access target areas, providing wider service coverage and faster service response~\cite{ref26,ref27}. With these advantages, UAVs can help the MEC system in two ways. On the one hand, UAVs can act as airborne MEC servers to provide air-ground cooperative computing services for TDs~\cite{ref30,ref31}. On the other hand, UAVs can act as repeaters to forward mission data between TDs and associated MEC servers~\cite{ref32,ref34,ref36}.\par

Despite the significant potential, the UAV-assisted MEC systems should consider both UAV trajectory design and limited onboard energy constraints, which are crucial for completing the interdependent sub-tasks. To date, some studies have explored dependency-aware sub-task offloading in UAV-assisted MEC scenarios~\cite{ref37,newDependenyUAVTask6,newDependenyUAVTask4,newDependenyUAVTask5,ref19,newDependenyUAVTask3,newDependenyUAVTask1}. Specifically, the authors in~\cite{ref37} investigated serial-dependency sub-task offloading in a single UAV scenario. Task dependencies were managed by imposing constraints on the start times of sub-tasks. To minimize system energy consumption, the study optimized sub-task offloading, computational resource allocation, and UAV trajectories. Similarly, the work in~\cite{newDependenyUAVTask6} also focused on a single UAV scenario, proposing a deep reinforcement learning algorithm that optimizes the UAV's trajectory to maximize the number of received dependent sub-tasks while adhering to limited energy constraints. For multi-UAV-assisted MEC scenarios, the authors of~\cite{newDependenyUAVTask4} explored an edge computing framework involving vehicle-UAV collaboration. They prioritized dependent sub-tasks based on transmission and computational costs, and then proposed a federated reinforcement learning-based task offloading scheme to minimize task execution latency and computational energy consumption. Additionally, the study in~\cite{newDependenyUAVTask5} proposed a task matrix to model sub-task dependencies and applied game theory to solve the joint optimization of task uplink assignments from TDs to UAVs and task transmission link assignments between UAVs. Further, the study in~\cite{ref19} considered not only the optimization of offloading decisions but also the optimization of communication resource allocation, and designed optimization algorithms based on Discrete Whale Optimization Algorithm (D-WOA) and Convex Optimization to minimize the average task completion delay. The research in~\cite{newDependenyUAVTask3} explored dependency-aware sub-task offloading and communication resource allocation in UAV-assisted smart farms, utilizing graph convolutional networks and reinforcement learning algorithms to minimize system energy consumption and task execution latency. While joint optimization of communication resources and sub-task offloading can reduce task completion latency and system energy consumption, optimizing computational resource allocation is critical in scenarios where computational demand exceeds communication demand. In this regard, the work in~\cite{newDependenyUAVTask1} proposed an improved constrained multiobjective evolutionary algorithm to minimize the task completion time as well as the difference in energy consumption between UAVs by optimizing sub-task scheduling and computational resource allocation. 

\subsection {Challenges in Task Offloading in Multi-UAV MEC}

In dealing with dependent sub-task offloading in a multi-UAV and multi-TD scenario, the key challenge is how to realize efficient sub-task scheduling and resource allocation among multiple UAVs in the presence of complex data transfer relationships among multiple sub-tasks, including one-to-one, many-to-one, and one-to-many dependency patterns. To ensure the dependencies among sub-tasks, existing studies usually prioritize the sub-tasks based on their execution time, or represent the timing and data constraints among sub-tasks by constructing sub-task dependency matrices. Based on this, the researchers further introduce the heuristic algorithm~\cite{ref19, newDependenyUAVTask1}, the reinforcement learning method~\cite{newDependenyUAVTask3,newDependenyUAVTask4}, and the game theory model~\cite{ newDependenyUAVTask5} to optimize the sub-task offloading decision and resource allocation scheme.\par

Most of the existing research is oriented to single-objective and multivariate optimization frameworks~\cite{ref19, newDependenyUAVTask5}, or multi-objective and univariate optimization frameworks~\cite{newDependenyUAVTask4}. however, considering the diversity of future application requirements, future research should form an optimization framework that simultaneously considers multiple objectives (e.g., simultaneously minimizing the task completion delay and system energy consumption), multiple constraints (e.g., task deadline delay and energy consumption upper limit), and multiple variables (e.g., offloading decision, computational and communication resource allocation, and UAV flight trajectory). Under such an optimization framework, multiple objective weights and variables can be flexibly adjusted to adapt to diverse scenarios.\par

Nevertheless, the highly coupled relationships among dependent sub-tasks and UAVs, along with the presence of multiple constraints, multiple objectives, and multiple variables, make the optimization problem exceedingly difficult to solve. On one hand, task dependencies strictly limit the scheduling order, on the other hand, computing and communication resource allocation greatly enlarge the search space. Further introducing UAV trajectory planning adds spatiotemporal complexity, which substantially raises the overall computational overhead. Existing methods for handling optimization problems involving multiple objectives, constraints, and variables often rely on heuristic or reinforcement learning algorithms~\cite{newDependenyUAVTask1,newDependenyUAVTask3}. however, as the decision space grows exponentially in larger-scale scenarios, these algorithms may fail to find sufficiently high-quality solutions. In particular, within a multiple-objective optimization setting, they tend to overemphasize one objective at the expense of others, thereby compromising overall system performance. Consequently, effective methods to address optimization problems involving multiple constraints, objectives, and variables for dependency-aware task offloading in large-scale multi-UAV and multi-TD environments remain lacking.\par

\subsection{Contribution}

To address the aforementioned issues, this article considers a multi-TD multi-UAV scenario, where each TD's task is divided into interdependent sub-tasks with either parallel or serial relationships. We propose a joint sub-task offloading, computational resource allocation and UAV trajectories scheme for multi-UAV assisted collaborative MEC systems. A two-timescale computing framework is proposed to handle sub-task chain dependencies. The joint optimization problem is formulated to minimize the system cost, defined as a weighted sum of task completion delay and system energy consumption. By adjusting the weight factors of the system cost, the formulation enables a tunable balance between latency- and energy-sensitive requirements, thereby improving the adaptability of the proposed scheme to diverse application scenarios. To solve this NP-hard mixed integer non-linear programming (MINLP) problem, we develop a Penalty Dual Decomposition and Successive Convex Approximation (PDD-SCA) algorithm, which can find the stationary solution. 

\begin{itemize}
\item We propose a novel multi-UAV multi-TD collaborative computing framework, which enables dependency-aware sub-task offloading. Specifically, the decomposed tasks of TDs are modeled into sub-task chains based on the DAG, which can be scheduled by a two-timescale frame structure. We formulate the joint sub-task offloading, computational resource allocation, and UAV trajectories problem to minimize system cost weighted by the task completion delay and the system energy consumption, which is an NP-hard MINLP problem.

\item We propose a dual-loop PDD-SCA algorithm to tackle the NP-hard MINLP problem. Particularly, we equivalently reformulate the original MINLP problem into a more tractable form based on PDD theory. Then, we decompose the transformed problem into three nested subproblems and employ the SCA method to recast the non-convex components into a convex form, thus obtaining the stationary solution.

\item We provide numerical results to evaluate the efficiency of the proposed PDD-SCA algorithm. Compared with the benchmark algorithms including the multiobjective evolutionary algorithm~\cite{newDependenyUAVTask1}, heuristic offloading algorithm, and fixed time/computational resource allocation algorithm, the proposed PDD-SCA algorithm can achieve lower system cost. Moreover, better workload balancing across UAVs can be realized even if the number of TDs increases.
\end{itemize}

The structure of the remaining part of this paper is as follows. Section II introduces the system model and problem formulation. Section III demonstrates the proposed PDD-SCA algorithm. Section IV presents the numerical results. Finally, conclusions are summarized in Section V.\par

\textit{Notations:} In this paper, we let bold upper-case letter $\bf A$, decorated letter $\cal A$, and italic letter $A$ or $a$ denote matrix, set, and scalar, respectively. Bold lower-case letter $\bf a$ denotes vector, $||{\bf a}||$ denotes
the Euclidean norm and $\bf A^\top$  denotes the transpose of $\bf A$. $\mathbb R^{M \times N}$ denotes the space of $M \times N$ real matrices.\par

\begin{figure}
    \centering
    \includegraphics[width=1\linewidth]{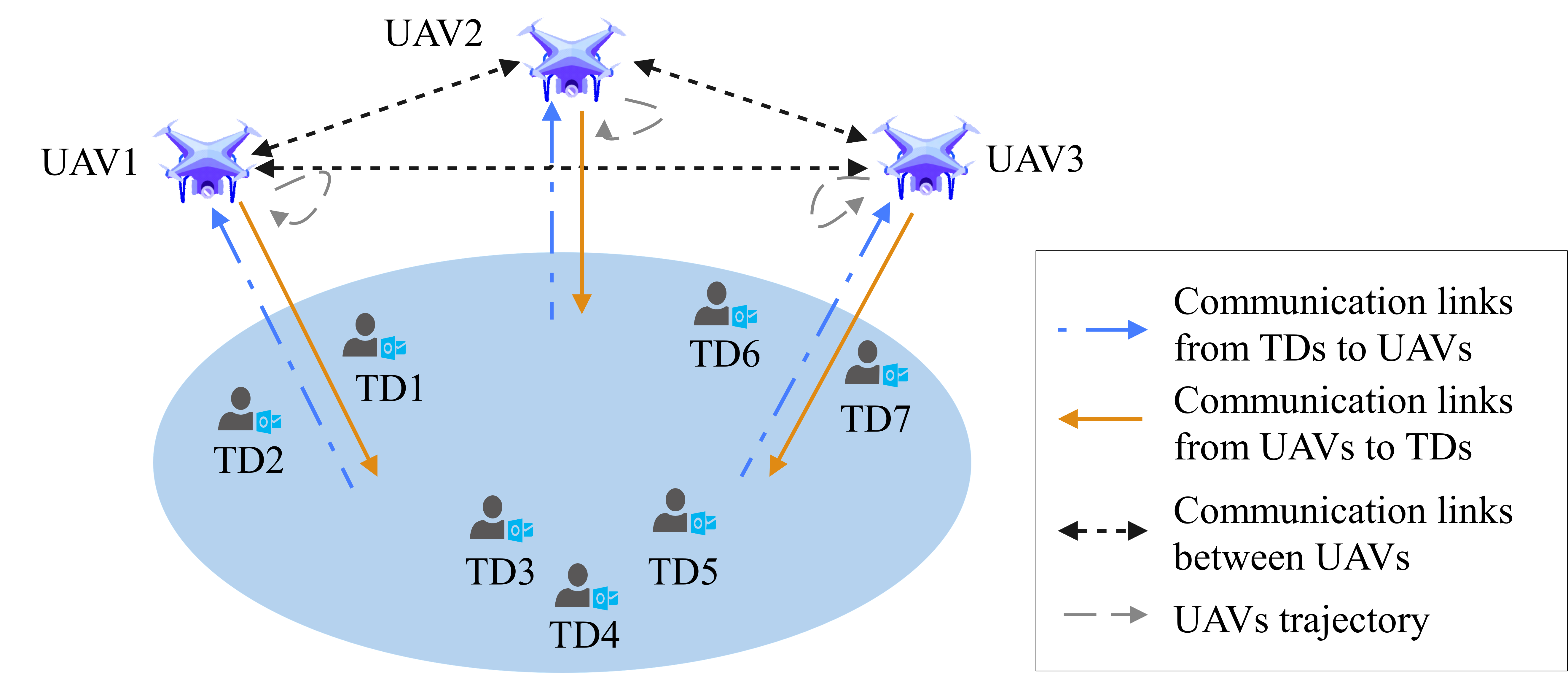}
    \caption{Multi-UAV assisted collaborative MEC system.}
    \label{fig1}
\end{figure}

\section{ System Model And Problem Formulation}
\subsection{System model}
As shown in Fig. \ref{fig1}, we consider a multi-UAV-assisted collaborative MEC system for remote or security areas, in which TDs cannot be connected to ground base stations. Each TD has a computationally intensive task, which can be subdivided into multiple sub-tasks that have dependencies on each other. The system involves $M$ UAVs with airborne MEC servers to provide services to $U$ TDs. The sets of $M$ UAVs and $U$ TDs are denoted by ${{\cal M}} = {\rm{\{ 1,}}...,M{\rm{\} }}$ and ${{\cal U}} = {\rm{\{ 1,}}...,U{\rm{\} }}$, respectively. Furthermore, let ${{\cal Z}} = \{ 1,...,M,M + 1,...,M + U\} $ denote the aggregate set of all devices (including both UAVs and TDs) in the system. \par
\begin{table}[!t]
\renewcommand{\arraystretch}{1.25}
\caption{System Parameters\label{tab:table1}}
\label{table1}
\centering
\begin{tabular}{|m{1.05cm}|m{6.85cm}|}
\hline
\textbf{Notation}& \textbf{Definition}\\ \hline
$\cal M$ & Set of UAVs, ${|\cal M|}= M$. \\ \hline
$\cal U$ & Set of TDs / tasks, ${|\cal U|}= U$. \\ \hline
$\cal Z$ & Set of all devices (including both UAVs and TDs), ${|\cal Z|}= M+U$. \\ \hline
$\cal N$ & Set of time slots, ${|\cal N|}= N$. \\ \hline
$\delta_n$ & Duration of time slot $n$. \\ \hline
$\tau^{\rm comm}[n]$ & Duration of communication unit of time slot $n$. \\ \hline
$\tau^{\rm comp}[n]$ & Duration of computation unit of time slot $n$. \\ \hline
$\cal T$ &  Set of communication and computation time units, ${|\cal T|}= 2N$. \\ \hline
${{\tilde \delta}_t}$ &  Duration of time unit $t$. \\ \hline
${\bf w}_u$ & Location of TD $u$. \\ \hline
${\bf q}_m[t]$ & Location of UAV $m$ at time slot $n$.  \\ \hline
$\bar B$ & Bandwidth of each sub-carrier. \\ \hline
$d_{mu}[t]$ & Distance between UAV $m$ and TD $u$ at time unit $t$. \\ \hline
$h_{mu}[t]$ & Channel gain between UAV $m$ and TD $u$ at time unit $t$. \\ \hline
$d_{mm'}[t]$ & Distance between UAV $m$ and UAV $m'$ at time unit $t$. \\ \hline
$h_{mm'}[t]$ & Channel gain between UAV $m$ and UAV $m'$ at time unit $t$. \\ \hline
$R_{zz'}[t]$ & Communication rate between device $z$ and device $z'$ at time unit $t$. \\ \hline
$l_u$ & Deadline of task $u$. \\ \hline
${\cal V}_u$ & Set of sub-task $v_u^k$, ${|{{\cal V}_u}|=K}$. \\ \hline
$v_u^k$ & The $k$-th sub-task of task $u$. \\ \hline
$c_u^k$ & The number of computation bits of sub-task $v_u^k$. \\ \hline
$o_u^{kk'}$ &  The number of transmission data bits from sub-task $v_u^k$ to $v_u^{k'}$.  \\ \hline
$x_{u,z}^{k}[n]$ & The sub-task offloading indicator for $v_u^k$ to the computing device $z$. \\ \hline
$v[n]$ & Set of sub-tasks that need to be computed in the time slot $n$. \\ \hline
$f_{u,z}^{k}[n]$ & The CPU frequency of device $z$ to execute $v_u^k$ in time slot $n$. \\ \hline
$E_{z}^{\rm comm}[n]$ & The communication energy consumption of device $z$ in time slot $n$. \\ \hline
$E_{z}^{\rm comp}[n]$ & The computation energy consumption of device $z$ in time slot $n$. \\ \hline
$P_{z}^{\rm prop}[t]$ & The propulsion power in time unit $t$ of UAV $z$. \\ \hline
$E_{z}^{\rm prop}[t]$ & The flight energy consumption of UAV $z$ in time unit $t$. \\ \hline
\end{tabular}
\end{table}

\begin{figure*}[ht]
    \centering
    \includegraphics[width=1 \linewidth]{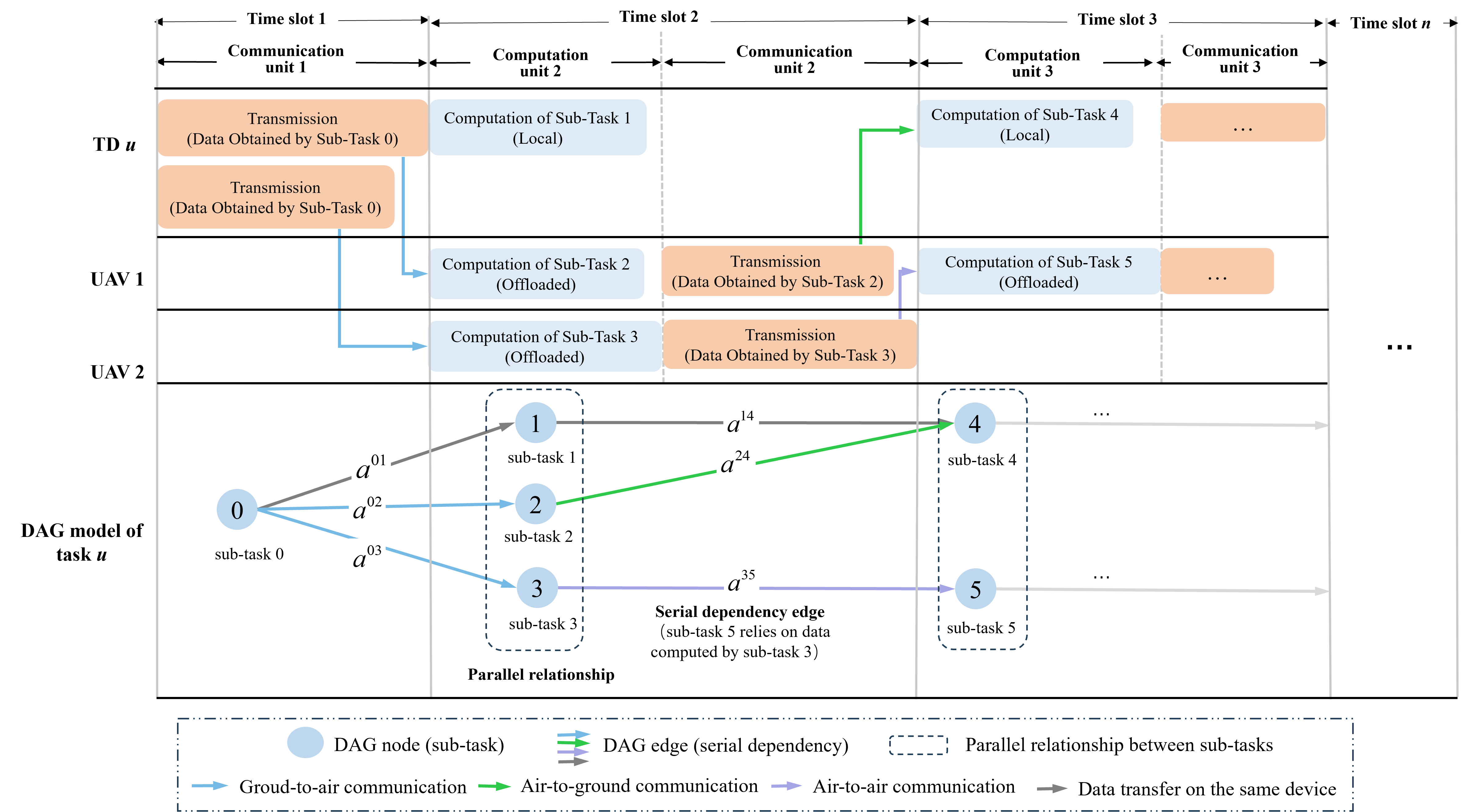}
    \caption{{An example of the two-timescale frame structure for dependency-aware sub-tasks computation and communication.}}
    \label{fig2}
\end{figure*}

The task of each TD can be decomposed into a set of interdependent sub-tasks. The set of sub-tasks needs to be completed during the UAV flight period $T$. We divide the period $T$ into $N$ slots, denoted as ${{\cal N}} = \{1, \ldots, N\}$. The duration of each time slot $n$ is denoted by ${\delta _n}$.  Furthermore, as shown in Fig. \ref{fig2}, each time slot $n$ is divided into two time units, namely the computation unit (for sub-task computing) and the communication unit (for data transfer between sub-tasks). The durations of these units are denoted by ${\tau ^{{\rm{comp}}}}[n]$ and  ${\tau ^{{\rm{comm}}}}[n]$, respectively.  As a result, the aggregate set of all these time units during the period $T$ can be represented as ${{\cal T}} = \{ 1,...,2N\} $. Let ${\tilde \delta _t}$ denote the duration of each time unit $t\in\mathcal{T}$, which yields the relationships ${\tau ^{{\rm{comp}}}}[n] = {\tilde \delta _{2n - 1}}$ and ${\tau ^{{\rm{comm}}}}[n] = {\tilde \delta _{2n}}$.\par

Without loss of generality, we assume that each TD $u$ has a fixed location in a 3D Cartesian coordinate system, with the horizontal location represented by $\mathbf{w}_{u} \in\mathbb{R}^{2\times 1}$ and the height set to $0$. \par

The amount of input data for a task is usually large, and to ensure complete transmission of the task input data, the UAV should remain stationary for the first time slot until the input data for the TD is fully received. Therefore, the length of the communication unit in the first time slot is not limited. For the communication units in the other time slots, the duration of the communication unit should be small enough in order for the UAV to remain stationary in each communication unit (the channel is considered constant). The duration of the remaining communication units, ${\tau ^{{\rm{comm}}}}[n]$, is constrained by $\tau _{\max }^{{\rm{comm}}}$, as expressed in
{\begin{equation}\label{eq1}
{\tau ^{{\rm{comm}}}}[n] \le \tau _{\max }^{{\rm{comm}}},\ \forall n \in \{2,3,...,N\}.
\end{equation}} \par

Let ${{\bf{q}}_m}[t] \in {\mathbb R}{^{2 \times 1}}$ and $H$ represent the horizontal coordinate and the fixed flight altitude of UAV $m$ at each communication/computation unit $t$, respectively. The UAV has a maximum flight speed of ${V_{\max }}$. Hence, the motion constraints of UAVs can be formulated as follows,
{\begin{align}
{||}{{\bf q}_m}[t + 1] - {{\bf q}_m}[t]|| \le {V_{\max }}{\tilde \delta _t},\ \forall m,t \label{eq2},\\
{||}{{\bf q}_m}[t] - {{\bf q}_{m'}}[t]|{|^2} \ge d_{\min }^2,\ \forall t,m,m' \ne m \label{eq3},
\end{align}}where ${d_{\min }}$ is the minimum secure distance to avoid collision between UAVs.\par

\textit{1) Task Model:} Each TD has a computationally intensive task that needs to be completed during period $\cal T$. For the sake of brevity, the set of tasks of TDs can also be expressed as ${{\cal U}}  = {\rm{\{ 1,}}...,U{\rm{\} }}$. The deadline of task $u$ is defined as ${l_u}$.

Each task $u$ can be decomposed into a chain of $K$ interdependent sub-tasks, indexed by ${{{\cal V}}_u}  = \{ v_u^1,...,v_u^K\} $. The number of computation bits of sub-task $v_u^k$ is denoted by $c_u^k$. \par

As shown in Fig. \ref{fig2}, we model both serial dependencies and parallel relationship among these sub-tasks based on a DAG model, which can be defined as $\mathcal{G}_{u}\left(\mathcal{V}_{u},\mathcal{A}_{u}\right)$. Specifically, ${{{\cal V}}_u}$ is the set of nodes in the DAG graph, i.e. the set of sub-tasks. ${{{\cal A}}_u}=\{ a_{u}^{kk'}|k\in {\cal K},k' \in {\rm{suc}}(v_u^k)\}$ denotes the set of edges between the nodes, where ${\rm{suc}}(v_u^k)$ denotes the set of succeeding sub-tasks of $v_{u}^{k}$. In detail, the edge $a_{u}^{kk'}$ from nodes $k$ to $k'$ represents the serial dependency between sub-tasks $v_{u}^{k}$ and $v_{u}^{k'}$, i.e., sub-task $v_{u}^{k'}$ can start computing only if sub-task $v_{u}^{k}$ is completed and the output data is transmitted. Define $o_u^{kk'}$ as the number of transmission data bits from $v_u^k$ to $v_u^{k'}$. On the other hand, two sub-tasks may be computed in parallel if they do not require computational data from each other, thus reducing the overall task completion latency. Furthermore, as shown in Fig. \ref{fig2}, at the beginning of each task, the initial input data will be transmitted from TD $u$ to the servers via sub-task $v_u^0$ at the first time slot. Similarly, at the last time slot, sub-task $v_u^{K+1}$ will be assigned for TD $u$ to receive the final output data from the servers. The number of computation bits of $v_u^0$ and $v_u^{K+1}$ are 0.\par

Based on the DAG task model, we propose a two-timescale frame structure that assigns sub-tasks to each time slot for computation and communication based on the order between sub-tasks and the time slot order, which ensures the dependencies between sub-tasks. As shown in Fig. \ref{fig2}, serial sub-tasks are completed within several consecutive time slots, while the parallel sub-tasks of each time slot $n$  are completed within their respective time slots. Define $v[n], \forall n \in {\cal N}$ as the set of parallel sub-tasks in time slot $n$. The duration of computation unit $\tau_{\rm comp}[n]$ is the maximum computation delay of the sub-tasks in $v[n]$, and the duration of communication unit $\tau_{\rm comm}[n]$ is the maximum communication delay of the sub-tasks in $v[n]$. The task $u$ is considered complete when the final sub-task $v_u^{K+1}$ has received all data.\par

\textit{2) Communication Model:} For the proposed multi-UAV assisted collaborative MEC system, the communication links include the uplink and downlink communications between TDs and UAVs, as well as the communication among UAVs. To avoid interference between data streams, orthogonal frequency division multiple access (OFDMA) is utilized in this system. The bandwidth of each sub-carrier is denoted by $\bar B = {{B(M + U)(M + U - 1)}/2}$, where $B$ represents the total bandwidth in the system.\par

The distance between the UAV $m$ and the TD $u$ at time unit $t$ is defined as ${d_{mu}}[t]=\sqrt {{H^2} + ||{{\bf q}_m}[t] - {{\bf w}_u}|{|^2}} $. Hence, the channel gain between UAV $m$ and TD $u$ can be expressed as ${h_{mu}}[t] = {{\beta _0}} /{d^2_{mu}}[t]$, where ${\beta _0}$ is the channel gain at a reference distance of $\rm 1\ m$. Similarly, the distance between two UAVs is expressed as ${d_{mm'}}[t] = \sqrt {||{{\bf q}_m}[t] - {{\bf q}_{m'}}[t]|{|^2}} $, and the corresponding channel gain between UAV $m$ and UAV $m'$ can be expressed as ${h_{mm'}}[t] = {{\beta _0}} /{d^2_{mm'}}[t]$. \par

\par

The uplink and downlink signal-to-noise ratios between the TD $u$ and the UAV $m$ in the time unit $t$ are denoted as ${\rm SNR}_{mu}^{\rm up}[t] = {{{p_u}{h_{mu}}[t]} /( {{\bar B}{N_0}}})$ and ${\rm SNR}_{mu}^{\rm down}[t] = {{{p_m}{h_{mu}}[t]} /( {{\bar B}{N_0}}})$, respectively, where ${p_u}{\rm{ = }}P_u^{\max }/M$ and ${p_m}{\rm{ = }}P_m^{\max }/(U + M - 1)$ represent the transmit power of TD $u$ and UAV $m$, with $P_{u}^{\max}$ and $P_m^{\max }$ representing the maximum transmit power of TDs and UAVs, respectively. In addition, $N_0$ represents the noise power spectral density. Correspondingly, the uplink and downlink communication rates between TD $u$ and UAV $m$ can be expressed as $R_{mu}^{\rm up}[t]$ and $R_{mu}^{\rm down}[t]$, respectively, with the following expressions
\begin{align}
    \label{eq4}
    R_{mu}^{\rm up}[t] &= \bar B{\log _2}\left( {1 + {\rm SNR}_{mu}^{\rm up}[t]} \right),\ \forall t,m,u, \\
    R_{mu}^{\rm down}[t] &= \bar B{\log _2}\left( {1 + {\rm SNR}_{mu}^{\rm down}[t]} \right),\ \forall t,m,u, \label{eq5}
\end{align}  Similarly, the signal-to-noise ratio between a UAV $m$ and a UAV $m'$ within a time unit $t$ is expressed as ${\rm SNR}_{mm'}^{\rm up}[t] = {{{p_m}{h_{mm'}}[t]}}/({{{\bar B}{N_0}}})$ and with the communication rate ${R_{mm'}}[t]$ can be expressed as follows,

\begin{equation}
{R_{mm'}}[t] = \left\{ {\begin{array}{*{20}{l}}
{\bar B{{\log }_2}\left( {1 + {\rm SNR}_{mm'}^{\rm up}[t]} \right),}&{\forall t, m \ne m',}\\
{\infty ,}&{\forall t, m = m'.}
\end{array}} \label{eq6} \right.
\end{equation}\par

Using the above definitions, we define the communication rate between any two devices during time unit $t$ as ${R_{zz'}}[t]$. For each $z \in \{ M + 1,...,M + U\}$ and $z' \in {{\cal M}}$, the rate ${R_{zz'}}[t] = R_{mu}^{\rm up}[t]$. Conversely, for each $z \in {{\cal M}}$ and $z' \in \{ M + 1,...,M + U\}$, the rate ${R_{zz'}}[t] = R_{mu}^{\rm down}[t]$. In cases where each $z$ and $z'$ are in ${{\cal M}}$ and $z \ne z'$, the rate is given as ${R_{zz'}}[t] = R_{mm'}[t]$.\par

\textit{3) Computing Model:} Note that the sub-tasks $v_u^0$ and $v_u^{K + 1}$ must be executed on the TD, while the other sub-tasks $v_{u}^{k}$, for $k \in \{1,...,K\}$, can be computed locally or offloaded to the UAVs. Let binary variable $x_{u,z}^k$ represent the sub-task offloading indicator for $v_{u}^{k}$ to the computing device $z$, where $z\in\{0,...,M\}$. Specifically, if $x_{u,0}^{k}=1$, the sub-task $v_{u}^{k}$  will be executed locally at TD $u$, otherwise, it is offloaded to a UAV. Based on the above definitions, It should be noted that in the rest of this paper, any instance of the device index $z=0$ should be interpreted as referring to the local computing device of TD $u$. \par

Since each sub-task can only be executed on one device, we have the following constraint
{\begin{equation} 
\sum\limits_{z = 0}^M {x_{u,z}^k}  = 1,\ \forall u, k.
\label{eq9}
\end{equation}}\par

The allocated CPU frequency of device to execute  $v_u^k$ in time slot $n$ is denoted as $f_{u,z}^k[n]$. Specifically, $f_{u,0}^k[n]$ is the allocated CPU frequency of TD $u$ to execute $v_u^k$, $f_{u,z}^k[n], \forall z \in {\cal M}$ is the allocated CPU frequency of UAV $m$ to execute $v_u^k$. In each time slot, the cumulative allocated CPU frequency of any device must not exceed its maximum CPU frequency, i.e., 
\begin{align}
&\sum\limits_k {f_{u,z}^k[n]}  \le F_u^{\max },\ \forall n, u, v_u^k \in v[n],z = 0,  \label{eq10} \\
&\sum\limits_u {\sum\limits_k {f_{u,z}^k[n]} }  \le F_m^{\max },\ \forall n, u, v_u^k \in v[n],z \in {\cal M},\label{eq11}
\end{align}where $F_u^{\max }$ and $F_m^{\max }$ represent the maximum CPU frequency of TDs and UAVs, respectively, and $v[n]$ is the set of parallel sub-tasks that need to be completed in the time slot $n$.\par

Let ${\vartheta _z}$ represent the required CPU cycles for computing each one bit of device $z$, the computation time of device $z$ in time slot $n$ is represented as
{\begin{align}
\tau _z^{{\rm{comp}}}[n] = \mathop {\max }\limits_{\forall v_u^k \in v[n]} \left( {\frac{{x_{u,z}^kc_u^k{\vartheta _z}}}{{f_{u,z}^k[n]}}} \right), \forall n, z \in \{0,...,M\}. \label{eq12}
\end{align}} \par

The duration of the communication unit $\tau _z^{\rm{comm}}[n]$ of the device $z$ is the maximum time for the associated sub-task on the device $z$ to transmit data to the succeeding dependent sub-task in time slot $n$, is given by
{\begin{align}
&\tau _z^{{\rm{comm}}}[n] = \mathop {\max }\limits_{\forall z' \ne z} \left( {\frac{{\sum\limits_u {\sum\limits_k {\sum\limits_{k'} {x_{u,z}^kx_{u,z'}^{k'}o_u^{kk'}} } } }}{{{R_{zz'}}[2n]}}} \right),\nonumber \\
&\forall n, v_u^k \in v[n],v_u^{k'} \in {\rm{suc}}(v_u^k), \forall z, z' \in \{ 0,...,M\}. \label{eq13}
\end{align}}\par

Furthermore, based on the aforementioned two-timescale frame structure, as shown in Fig. \ref{fig2}, the duration of each time unit can be dynamically allocated. Therefore, the parallel sub-task completion latency of each time slot $n$, i.e., the duration of time slot $\delta_n$, is equal to the sum of the maximum computation unit duration and the maximum communication unit duration within time slot $n$, and is expressed as
{\begin{equation}
{\delta _n}{\rm{ = }}\mathop {\max }\limits_{\forall z} \left( {\tau _z^{{\rm{comp}}}[n]} \right) + \mathop {\max }\limits_{\forall z} \left( {\tau _z^{{\rm{comm}}}[n]} \right), \forall n \in {\cal N}. \label{eq14}
\end{equation}} \par

Each task has a deadline of $l_u$, thus we have
{\begin{equation}
\sum\limits_{n = 1}^N {{\delta _n}}  \le \mathop {\min }\limits_{\forall u} \left( {{l_u}} \right). \label{eq15}
\end{equation}}

\textit{4) Energy Consumption Model:} The communication energy consumption of device $z$ in time slot $n$, denoted by $E_z^{{\rm{comm}}}[n]$, is expressed as
{\begin{align}
&E_z^{{\rm{comm}}}[n] = \sum\limits_{\forall z' \ne z} {{p_z}\left( {\frac{{\sum\limits_u {\sum\limits_k {\sum\limits_{k'} {x_{u,z}^kx_{u,z'}^{k'}o_u^{kk'}} } } }}{{{R_{zz'}}[2n]}}} \right)} ,\nonumber \\
&\forall n, v_u^k \in v[n],v_u^{k'} \in {\rm{suc}}(v_u^k),\forall z,z' \in \{ 0,..,M\}, \label{eq16} 
\end{align}}where $p_z$ represents the transmit power of device $z$.\par

The computation energy consumption of device $z$ in time slot $n$, denoted by $E_z^{{\rm{comp}}}[n]$, is expressed as
{\begin{align}
&E_z^{{\rm{comp}}}[n] = \sum\limits_u {\sum\limits_k {x_{u,z}^kc_u^k{\vartheta _z}{\gamma _z}{{(f_{u,z}^k[n])}^2}} } ,\nonumber \\
&\forall n, v_u^k \in v[n],\forall z \in \{ 0,..,M\},   \label{eq17}
\end{align}}where ${\gamma _z}$ denotes the effective capacitance coefficient affected by chip architecture at device $z$.\par

For UAV flight energy consumption, the speed of UAV $m$ in time unit $t$ can be given by
\begin{equation}
{{\bf v}_m}[t] = \frac{{{{\bf q}_m}[t + 1] - {{\bf q}_m}[t]}}{{{{\tilde \delta }_t}}},\forall t, m\in {\cal M}. \label{eq18}
\end{equation}\par

Based on the~\cite{ref32,ref39}, the propulsion power in each time unit $t$ of UAV $m$ is expressed as
\begin{align}
P_m^{{\rm{prop}}}[t] = {P_0}\left( {1 + \frac{{3||{{\bf v}_m}[t]|{|^2}}}{{U_{tip}^2}}} \right) + \frac{1}{2}{d_0}\beta As||{{\bf v}_m}[t]|{|^3}\nonumber \\
\ \ \  + {P_i}{\left( {\sqrt {1 + \frac{{||{{\bf v}_m}[t]|{|^4}}}{{4v_0^2}}}  - \frac{{||{{\bf v}_m}[t]|{|^2}}}{{2v_0^2}}} \right)^{\frac{1}{2}}}, \forall t, m\in {\cal M},  \label{eq19}
\end{align}where $P_0$ and $P_i$ represent the blade profile power and induced power in hovering status, respectively. The other parameters of $U_{\rm tip}$, $v_0$, $d_0$, $\beta$, $s$, and $A$ related to the UAV’s aerodynamics.\par

The flight energy consumption of UAV $m$ in time unit $t$ is expressed as 
{\setlength{\jot}{2pt}{
\begin{equation}
E_m^{{\rm{prop}}}[t] = {\tilde \delta _t}P_m^{{\rm{prop}}}[t], \forall t, m\in {\cal M}. \label{eq20}
\end{equation} Then the total flight energy consumption in time slot $n$ of UAV $m$ is obtained as
\begin{equation}
E_m^{{\rm{prop}}}[n] = E_m^{{\rm{prop}}}[2n - 1] + E_m^{{\rm{prop}}}[2n], \forall n, m\in {\cal M}.
\label{eq21}
\end{equation}}\par

The energy consumption for communication, computation, and UAV flight are limited by the following constraints 
{\setlength{\jot}{4pt}
{\begin{align}
\sum\limits_{n = 1}^N {\left( {E_z^{{\rm{comm}}}[n] + E_z^{{\rm{comp}}}[n]} \right)}  \le E_{z,{\rm{com}}}^{{\rm{max}}},\ \forall{\rm{ }}z, \label{eq22} \vspace{0.5ex}\\
\sum\limits_{n = 1}^N {E_m^{{\rm{prop}}}[n]}  \le E_{m,{\rm{prop}}}^{{\rm{max}}},\ \forall m,\label{eq23}
\end{align}}}where $E_{z,{\rm{com}}}^{{\rm{max}}}$ represents the maximum energy consumption for communication and computation of the device $z$ (TD or UAV), and $E_{m,{\rm{prop}}}^{{\rm{max}}}$ indicates the energy consumption limit for UAV flight. \par

\subsection{Problem Formulation}

In multi-UAV-assisted MEC systems, both task completion delay and energy consumption are critical performance metrics. We aim to optimize these objectives simultaneously to improve overall system efficiency. We use the weighted-sum method~\cite{weightedSumMOO} to scalarize the multi-objective problem into a single-objective formulation. Specifically, we introduce the sub-task offloading variables ${\bf X} = \{ x_{u,z}^k,\ \forall u, k, z\}$, the computational resource allocation ${\bf F} = \{ f_{u,z}^k[n],\ \forall u, k, n, z\}$, and the UAV trajectories ${\bf Q} = \{ {\bf q}_m[t],\ {\bf v}_m[t],\ \forall t, m\}$. Then, we define the optimization objective as the weighted sum of task completion delay and system energy consumption, which we refer to as the system cost, expressed as
\begingroup
\begin{align}
\Omega({\bf X}, {\bf F}, {\bf Q}) &= w^{\rm tim}\sum_{n}\delta_n
+ \sum_{z} \sum_{n} w_z^{\rm com} ( E_z^{\rm comm}[n] +  \nonumber \\
&\quad E_z^{\rm comp}[n] ) + \sum_{m} \sum_{n} w_m^{\rm fly} E_m^{\rm prop}[n],
\label{syscosFun}
\end{align}\endgroup where $w^{\rm tim}$ denotes the weight factor of task completion delay, $w_z^{{\rm{com}}}, \forall z \in \cal Z$ denotes the weight factor of computation and communication energy consumption, $w_m^{{\rm{fly}}}, \forall m \in \cal M$ denotes the weight factor of UAV flight energy consumption. In specific scenarios, the weight factors can be flexibly adjusted to shift the optimization focus according to the desired performance trade-off. \par

Based on the above discussion, the system cost minimization problem is formulated as
\begin{subequations} \label{P1}
\begin{align} 
 {\rm{(P1):}} &\mathop {\min }\limits_{\{ \bf X,\bf F,\bf Q\} } \Omega(\bf X,\bf F,\bf Q) 
\nonumber \\ 
{\rm{s.t.}}\ &{ \eqref{eq1}-\eqref{eq3},\ \eqref{eq9}-\eqref{eq11},\ \eqref{eq15},\ \eqref{eq22},\ \eqref{eq23},} \nonumber \label{eq24a} \\ 
&{{\bf q}_m}[0] = {{\bf q}_m}[2N + 1],\ {\rm{   }}\forall m,   \\ \label{eq24b}
& x^k_{u,z} \in \{0,1\},\  \forall u, k, z \in \{0,...,M\}, 
\end{align}
\end{subequations} where constraint \eqref{eq1} denotes the maximum duration of communication units. Constraint \eqref{eq2} indicates the maximum flight distance of UAVs in each time unit, and constraint \eqref{eq3} denotes the minimum secure distance for any two UAVs for collision avoidance.  Constraint \eqref{eq9} means that each sub-task can be assigned to only one device.     Constraints \eqref{eq10} and \eqref{eq11} indicate the maximum computational resource for devices.    Constraint \eqref{eq15} denotes the deadline requirements of the task completion delay.    Constraints \eqref{eq22} and \eqref{eq23} represent the energy consumption limitations of the system. Constraint \eqref{eq24a} indicates that the UAV needs to fly back to its original location to facilitate charging if the battery power is not sufficient.  Constraint \eqref{eq24b} is the binary constraint for sub-task offloading.\par

(\hyperref[P1]{P1}) involves the sub-task offloading optimization, which is analogous to a Generalized Assignment Problem (GAP),  widely recognized as an NP-hard problem~\cite{ref19, GAP-NP}. Therefore, (\hyperref[P1]{P1}) is an NP-hard MINLP problem, for which finding a globally optimal solution is challenging. An efficient suboptimal solution algorithm will be proposed in the following section. 

\section{PDD-SCA Algorithm Based Joint Optimization}
To solve (\hyperref[P1]{P1}), we propose a dual-loop PDD-SCA algorithm based on the penalty dual decomposition (PDD) and successive convex approximation (SCA) algorithms. The penalty factor and augmented lagrangian (AL) multipliers are updated in the outer loop, while the auxiliary variables, sub-task offloading, computational resource allocation, and UAV trajectories are optimized in the inner loop.

\subsection{Integer Variable Relaxations}
According to the PDD framework in~\cite{ref38}, we introduce auxiliary variables $\{ \tilde x_{u,z}^k,\forall u, k, z \in \{ 0,...,M\} \} $, then recast constraint \eqref{eq24b} as
{\begin{align}
x_{u,z}^k(\tilde x_{u,z}^k - 1) = 0,{\rm{   }}\forall u, k,z \in \{ 0,...,M\}, \label{eq25} \\
x_{u,z}^k - \tilde x_{u,z}^k = 0,{\rm{   }}\forall u, k, z \in \{ 0,...,M\}, \label{eq26} \\
0 \le x_{u,z}^k \le 1,{\rm{   }}\forall u, k, z \in \{ 0,...,M\}, \label{eq27} \\
0 \le \tilde x_{u,z}^k \le 1,{\rm{   }}\forall u, k, z \in \{ 0,...,M\}.  \label{eq28} 
\end{align}}\par

We further define ${\bf \tilde X = }\{ \tilde x_{u,z}^k,\forall u, k, z \in {\rm{\{ 0,}}...{\rm{,}}M{\rm{\} }}\}$, and $\Theta = \{ \bf \tilde X,X,F,Q \}$. By dualizing and penalizing the equality constraints into the objective function as AL items, (\hyperref[P1]{P1}) can be transformed into (\hyperref[P2]{P2}),
{\begin{subequations}
\label{P2}
\begin{align} 
& {\rm{(P2):   }}\mathop {\min }\limits_{\Theta } \Omega( \Theta) + \frac{1}{{2{\varrho }}}\sum\limits_u {\sum\limits_k {\sum\limits_z {\psi _{u,z}^k} } }  \nonumber \\
&{\rm{s.t.}}\ { \eqref{eq1}-\eqref{eq3},\eqref{eq9}-\eqref{eq11},\eqref{eq15},\eqref{eq22},\eqref{eq23},\eqref{eq24a},\eqref{eq27},\eqref{eq28} }, \nonumber 
\end{align}
\end{subequations}}where $\psi _{u,z}^k = { || {x_{u,z}^k(\tilde x_{u,z}^k - 1) + {\varrho }\lambda _{u,k,z}^1} ||^2} + {|| {x_{u,z}^k - \tilde x_{u,z}^k + {\varrho }\lambda _{u,k,z}^2} ||^2}$ is the AL term, $\{ \lambda _{u,k,z}^1,\lambda _{u,k,z}^2\},\ \forall u, k, z \in \{ 0,...,M\}$ denote the AL multipliers, $\varrho  \in {{\mathbb R}_ + }$ represents the non-negative penalty parameter. The AL multiplier or the penalty factor is updated according to the value of the constraint violation indicator to guarantee the binary constraints, which is described in section IV-C. When $\varrho  \to 0$, (\hyperref[P2]{P2}) is consistent with (\hyperref[P1]{P1}). Based on the discussion of~\cite{ref38} and~\cite{ref40}, the PDD-based algorithm has a guaranteed convergence to a KKT point of the problem. \par

\subsection{Inner-Loop Update of PDD-SCA Algorithm}
In (\hyperref[P2]{P2}), to make the problem more tractable, we divided $\Theta$ into three blocks, which are ${\Theta _1} = \{ {{\bf \tilde X}} \}$, ${\Theta _2} = \left\{ {{\bf X,F}} \right\}$, and ${\Theta _3} = \left\{ {{\bf Q} } \right\}$.\par

\textit{1) Auxiliary variables optimization:} To optimize block ${\Theta _1} = \{ \tilde {\mathbf X}\}$, given the values of ${\Theta _2}$ and ${\Theta _3}$, the sub-problem is formulated as
\begin{equation}
 \label{P3}
 {\rm{(P3):}}\mathop {\min }\limits_{\{ {\Theta_1 }\} } \frac{1}{{2\varrho }}\sum\limits_u {\sum\limits_k {\sum\limits_z \psi _{u,z}^k } } \nonumber
\end{equation}
\par
\setcounter{equation}{27}
Clearly, (\hyperref[P3]{P3}) is a convex problem that is easy to solve. The closed-form solution for problem (\hyperref[P3]{P3}) can be obtained by directly computing its derivative as follows,
{\setlength{\jot}{0pt}
{\begin{align}
&\tilde x_{u,z}^k = \frac{{{{\left( {x_{u,z}^k} \right)}^2} + x_{u,z}^k - \varrho \lambda _{u,k,z}^1x_{u,z}^k + \varrho \lambda _{u,k,z}^2}}{{{{\left( {x_{u,z}^k} \right)}^2} + 1}},{\rm{  }}\nonumber \\
&\forall u, k, z \in \{ 0,...,M\}.  \label{eq29}
\end{align}}}\par

\textit{2) Sub-task offloading and computational resource optimization:} First, we introduce slack variables ${\tau _{1,z}}[n]$, ${\tau _{2,z,z'}}[n]$, ${\tau _1}[n]$, and ${\tau _2}[n]$ as follows,
{\setlength{\jot}{0pt}
{\begin{align}
&{\tau _{1,z}}[n]\ge \mathop {\max }\limits_{\forall v_u^k \in v[n]} \left( {\frac{{x_{u,z}^kc_u^k{\vartheta _z}}}{{f_{u,z}^k[n]}}} \right), \forall n, u, z, \label{eq30} \\
&{\tau _{2,z,z'}}[n]\ge \frac{{\sum\limits_u {\sum\limits_k {\sum\limits_{k'} {x_{u,z}^kx_{u,z'}^{k'}o_u^{kk'}} } } }}{{{R_{zz'}}[2n - 1]}},{\rm{  }} \nonumber \\
&\forall n,z,z', v_u^k \in v[n],v_u^{k'} \in {\rm{suc}}(v_u^k), \label{eq31} \\
&{\tau _1}[n] \ge \mathop {\max }\limits_{\forall z} \left( {{\tau _{1,z}}[n]} \right), \forall n, \label{eq32}\\
&{\tau _2}[n]\ge \mathop {\max }\limits_{\forall z} \left( {\mathop {\max }\limits_{\forall z'} \left( {{\tau _{2,z,z'}}[n]} \right)} \right), \forall n. \label{eq33}
\end{align}}}\par

Next, we define ${d_m}[t] = ||{{\bf q}_m}[t + 1] - {{\bf q}_m}[t]||,\forall t \in {{\cal T}}$. The corresponding scalar speed can be expressed as ${{v}_m}[t] = {d_m}[t]/{\tilde \delta _t}$, where ${\tilde \delta _{2n - 1}} = {\tau _1}[n]$ and ${\tilde \delta _{2n}} = {\tau _2}[n]$.\par

According to the above definitions, Eq. \eqref{eq20} can be re-expressed as
{\begin{align}
E_m^{{\rm{prop}}}[t] &= {P_0}\left( {{{\tilde \delta }_t} + \frac{{3{d_m}{{[t]}^2}}}{{{{\tilde \delta }_t}U_{tip}^2}}} \right) + \frac{1}{2}{d_0}\beta As\frac{{{d_m}{{[t]}^3}}}{{{{\tilde \delta }_t}^2}} \nonumber\\
 &+ {P_i}{\left( {\sqrt {{{\tilde \delta }_t}^4 + \frac{{{d_m}{{[t]}^4}}}{{4v_0^2}}}  - \frac{{{d_m}{{[t]}^2}}}{{2v_0^2}}} \right)^{\frac{1}{2}}}, \forall t, m. \label{eq34}
\end{align}}Similarly, Eq. \eqref{eq22} can be re-expressed as
\vspace{0pt} 
\begin{flalign}
&\sum\limits_n {\sum\limits_u {\sum\limits_k {x_{u,z}^kc_u^k{\vartheta _z}{\gamma _z}{{(f_{u,z}^k[n])}^2}} } }  + \sum\limits_{z'} {{p_z}{\tau _{2,z,z'}}[n]}  \le E_{z,\max }^{{\rm{com}}},{\rm{ }} \nonumber \\
&\forall z, z', v_u^k \in v[n],v_u^{k'} \in {\rm{suc}}(v_u^k)  \label{eq35c}. 
\end{flalign}\par

Based on the above discussion, define $\Theta'_2 = \{ \Theta_2,\ \left\{ \tau_{1,z}[n] \right\}_{n,z},\ \left\{ \tau_{2,z,z'}[n] \right\}_{n,z,z'},\ \left\{ \tau_1[n] \right\}_n,\ \left\{ \tau_2[n] \}_n \right\}
$, the subproblem of optimizing $\Theta'_2$ with the given values of ${\Theta _1}$ and ${\Theta _3}$ can be expressed as\par
\vspace{-8pt} 
\begingroup
\setlength{\abovedisplayskip}{1pt}  
\setlength{\belowdisplayskip}{1pt}
\begin{subequations} \label{P4}
\begin{align}
{\rm{(P4):    }}&\mathop {\min } \limits_{ \Theta'_2} {\rm{ }}\Omega( \Theta'_2) {\rm{ + }}\frac{1}{{2\varrho }}\sum\limits_u {\sum\limits_k {\sum\limits_z { {\psi _{u,z}^k} } } } \nonumber \\
{\rm{s.t.}}\ &{ \eqref{eq9}-\eqref{eq11},\ \eqref{eq23},\ \eqref{eq27},\ \eqref{eq30}-\eqref{eq33} },\  \eqref{eq35c},\nonumber \\
&{\tau _2}[n] \le \tau _{\max }^{{\rm{comm}}},{\rm{    }}\ \forall n \in \{2,3,...,N\}, \label{eq35a} \\
&\sum\limits_{n = 1}^N {\left( {{\tau _1}[n] + {\tau _2}[n]} \right)}  \le \mathop {\min }\limits_{\forall u \in {{\cal U}}} \left( {{l_u}} \right),  \label{eq35b}\\
&{{ v}_m}[t] \le {V_{\max }}, \forall t, m. \label{eq35d}
\end{align}
\end{subequations}
\endgroup
\setcounter{equation}{35}

In (\hyperref[P4]{P4}), besides the non-convex objective function, constraints \eqref{eq23}, \eqref{eq30}, \eqref{eq31}, and \eqref{eq35c} are also non-convex. It can be observed that in constraints \eqref{eq30}, \eqref{eq31}, and \eqref{eq35c}, the non-convexity mainly results from three types of structural expressions, i.e.,  $x_{u,z}^kx_{u,z'}^{k'}$, ${{x_{u,z}^k}}/f_{u,z}^k[n]$, and $x_{u,z}^k{(f_{u,z}^k[n])^2}$. For ${{x_{u,z}^k}}/f_{u,z}^k[n]$ and $x_{u,z}^k{(f_{u,z}^k[n])^2}$, we introduce slack variables as 
{\begin{align}
& {\widehat f} _{u,z}^k[n] \ge {1 / {f_{u,z}^k[n]}}, \forall v_u^k \in v[n],\forall n,z, \label{1/f} \\
& \hat f_{u,z}^k[n] \ge {\left( {f_{u,z}^k[n]} \right)^2}, \forall v_u^k \in v[n],\forall n,z, \label{f^2} \end{align}}then we can get $x_{u,z}^k{\widehat f} _{u,z}^k[n]$, and $x_{u,z}^k\hat f_{u,z}^k[n] $. To tackle the non-convexity of $x_{u,z}^kx_{u,z'}^{k'}$, $x_{u,z}^k{\widehat f} _{u,z}^k[n]$, and $x_{u,z}^k\hat f_{u,z}^k[n] $, the SCA technique is applied.\par

First, for $x_{u,z}^kx_{u,z'}^{k'}$, we can convert it to
\begin{equation}
x_{u,z}^kx_{u,z'}^{k'} = {{{{( {x_{u,z}^k + x_{u,z'}^{k'}} )}^2}} \over 2} - {{{{( {x_{u,z}^k} )}^2}} \over 2} - {{{{( {x_{u,z'}^{k'}} )}^2}} \over 2}, \label{eq36}
\end{equation} then for given any local points $\{x_{u,z}^{k,j},\ x_{u,z'}^{k',j}\}$ ($j$ denotes $j$-th iteration), ${{{\left( {x_{u,z}^k} \right)}^2}} / 2$ and ${{{{( {x_{u,z'}^{k'}} )}^2}} / 2}$ can be lower-bounded via the first-order Taylor expansion as it is jointly convex with respect to $\{x_{u,z}^{k},\ x_{u,z'}^{k'}\}$. Let $({x_{u,z}^kx_{u,z'}^{k'}})_{\rm appro}$ denote the approximate function of $x_{u,z}^kx_{u,z'}^{k'}$, which is expressed as
{\setlength{\jot}{2pt}
\begin{align}
({x_{u,z}^kx_{u,z'}^{k'}})_{\rm appro} &= {{{{( {x_{u,z}^k + x_{u,z'}^{k'}} )}^2}} \over 2} - {{2x_{u,z}^kx_{u,z}^{k,j} - {{( {x_{u,z}^{k,j}} )}^2}} \over 2}  \nonumber \\
  &  - {{2x_{u,z'}^{k'}x_{u,z'}^{k',j} - {{( {x_{u,z'}^{k',j}} )}^2}} \over 2}, \label{eq37}
\end{align}}  \par

\renewcommand{\baselinestretch}{1} Similar to \eqref{eq37}, $x_{u,z}^k{\widehat f}_{u,z}^{k}$, and $x_{u,z}^k{\hat f}_{u,z}^{k}$ can obtain their approximate functions as follows,
{\begin{align}
(x_{u,z}^k{\widehat f}_{u,z}^{k}) _{\rm appro} & = \frac{{{{( {x_{u,z}^k + {\widehat f} _{u,z}^k[n]} )}^2}}}{2} - \frac{{2x_{u,z}^kx_{u,z}^{k,j} - {{( {x_{u,z}^{k,j}} )}^2}}}{2} \nonumber \\
&- \frac{2{\widehat f} _{u,z}^k[n]{\widehat f} _{u,z}^{k,j}[n] - ({\widehat f} _{u,z}^{k,j}[n] )^2}{2}, \label{eq38}\\
(x_{u,z}^k{\hat f}_{u,z}^{k})_{\rm appro} & = \frac{{{{( {x_{u,z}^k + \hat f_{u,z}^k[n]} )}^2}}}{2} - \frac{{2x_{u,z}^kx_{u,z}^{k,j} - {{( {x_{u,z}^{k,j}} )}^2}}}{2}\nonumber \\
& - \frac{{2\hat f_{u,z}^k[n]\hat f_{u,z}^{k,j}[n] - {{( {\hat f_{u,z}^{k,j}[n]} )}^2}}}{2}, \label{eq39}
\end{align}} \par

\renewcommand{\baselinestretch}{1}Next, we address the non-convexity of the objective function of (\hyperref[P4]{P4}) and constraint \eqref{eq23}. To handle the non-convex term ${P_i}{\left( {\sqrt {{{\tilde \delta }_t}^4 + \frac{{{d_m}{{[t]}^4}}}{{4v_0^2}}}  - \frac{{{d_m}{{[t]}^2}}}{{2v_0^2}}} \right)^{\frac{1}{2}}}$ of Eq. \eqref{eq34}, we introduce slack variable ${u_m}{[t]^2} \ge \sqrt {{{\tilde \delta }_t}^4 + \frac{{{d_m}{{[t]}^4}}}{{4v_0^2}}}  - \frac{{{d_m}{{[t]}^2}}}{{2v_0^2}}$, which can be simplified to ${\tilde \delta _t}^4 \le {u_m}{[t]^4} + \frac{{{d_m}{{[t]}^2}{u_m}{{[t]}^2}}}{{v_0^2}}$.  Since ${u_m}{[t]^4} + \frac{{{d_m}{{[t]}^2}{u_m}{{[t]}^2}}}{{v_0^2}}$ is convex with respect to $u_m[t]$, by given local points $u_m^j[t]$ ($j$ represents $j$-th iteration), we can apply Taylor's first-order expansion to it and obtain
\vspace{-1pt} 
{\begin{align}
{{\tilde \delta }_{t}}^4& \le ({u_m}[t] - u_m^j[t])(4u_m^j{[t]^3} + \frac{2{{d_m}{{[t]}^2}u_m^j[t]}}{{v_0^2}}) \nonumber \\
& + \frac{{{d_m}{{[t]}^2}u_m^j{{[t]}^2}}}{{v_0^2}} + u_m^j{[t]^4}. \label{eq40}
\end{align}}
\vspace{-3pt} 
Building on the previous discussion, substituting \eqref{eq38}, \eqref{eq37}, \eqref{1/f}, \eqref{f^2} into \eqref{eq30}, \eqref{eq31}, we can get ${\tau _{1,z}}[n] \ge \mathop {\max }\limits_{\forall v_u^k \in v[n]} ({(x_{u,z}^k\hat f_{u,z}^k[n])_{{\rm{appro}}}}$$c_u^k{\vartheta _z})$ and ${\tau _{2,z,z'}}[n] \ge \sum\limits_u {\sum\limits_k {\sum\limits_{k'} {(x_{u,z}^k} } }$$x_{u,z'}^{k'}{)_{{\rm{appro}}}}o_u^{kk'}/{R_{zz'}}[2n - 1]$. Then, $\delta_n$, $E_z^{\rm comm}[n]$, $E_m^{\rm comp}[n]$, and $E_z^{\rm prop}[t]$ can be re-expressed as approximate convex functions as
{\setlength\abovedisplayskip{2pt}
{\setlength\belowdisplayskip{2pt}
{\begin{align}
& {\delta _{n,{\rm{appro}}}} = {\tau _1}[n] + {\tau _2}[n], \forall n \in {\cal N}, \label{newDelataN} \\
& E_{z,{\rm{appro}}}^{{\rm{comm}}}[n] = p_z{\tau _2}[n], \forall n \in {\cal N}, z \in {\cal Z}, \label{newEcomm} \\
& E_{z,{\rm{appro}}}^{{\rm{comp}}}[n] = {(x_{u,z}^k\hat f_{u,z}^k[n])_{{\rm{appro}}}}c_u^k{\vartheta _z}{\gamma _z}, \nonumber \\
& \ \ \ \ \ \ \ \ \ \ \ \ \  \forall n \in {\cal N}, z \in {\cal Z}, \label{newEcomp} \\
& E_{m,{\rm{appro}}}^{{\rm{prop}}}[t] = {P_0}( {{{\tilde \delta }_t} + {{3{d_m}{{[t]}^2}} / ({{{\tilde \delta }_t}U_{tip}^2})}} ) + {1 \over 2}{d_0}\beta As{{{d_m}{{[t]}^3}} / {{{\tilde \delta }_t}^2}}  \nonumber \\
&\ \ \ \ \ \ \ \ \ \ \ \ \  + P_i{u_m}{[t]}, \forall n \in {\cal N}, m \in {\cal M}.  \label{newEprop}
\end{align}}}}\par
\vspace{-3pt} 
From Eq. \eqref{newDelataN}-Eq. \eqref{newEcomp}, it can be observed that in the system cost, when the weighted task completion delay is greater than the weighted system energy consumption, the system will reduce the computation delay and the communication delay by increasing the allocated computational resources as well as by assigning dependent sub-tasks to the same device. Conversely, when the weighted system energy consumption is greater than the weighted task completion delay, the system will reduce the computational resources associated with sub-tasks to reduce computational energy consumption.
\vspace{-3pt} 
Define $\Theta''_{2} = \{\Theta'_2,\ \{ \widehat{f}_{u,z}^t[n] \}_{n,z,t},\{ \hat{f}_{u,z}^t[n] \}_{n,z,t},$ $\{ u_m[t] \}_{m,t} \}$. By substituting the approximately convex expressions into Eq.~\eqref{syscosFun}, the system cost function can be reformulated as
{\begin{align}
    \Omega(\Theta''_2)=  &
w^{\rm tim}\sum_{n} \delta_{n,\mathrm{appro}} 
+ \sum_{z} \sum_{n} w_z^{\mathrm{com}} ( E_{z,\mathrm{appro}}^{\mathrm{comm}}[n] + \nonumber \\ 
 & E_{z,\mathrm{appro}}^{\mathrm{comp}}[n] ) 
+ \sum_{m} \sum_{t} w_m^{\mathrm{fly}} E_{m,\mathrm{appro}}^{\mathrm{prop}}[t] .
\end{align}
Then, based on the above reformulation, the original problem (\hyperref[P4]{P4}) can be transformed into a convex problem (\hyperref[P4.1]{P4.1}), which can be efficiently solved using standard convex optimization solvers such as CVX~\cite{ref41}.
\vspace{-1pt} 
\begin{subequations} 
\label{P4.1}
\begin{align}
&{\rm{(P4.1):}}\mathop {\min } \limits_{\{ \Theta''_{2}\} } \Omega(\Theta''_{2}) + \frac{1}{2\varrho} \sum_u \sum_k \sum_z \psi_{u,z}^k \nonumber \\
&{\rm{s.t.}}\ 
\eqref{eq9}–\eqref{eq11},\ \eqref{eq27},\ \eqref{eq30}–\eqref{eq33},\ \eqref{eq35a},\ \eqref{eq35b},\ \eqref{eq35d},\nonumber \\
&\eqref{1/f},\ \eqref{f^2},\ \eqref{eq40}, \nonumber \\
&\sum_{n = 1}^N \left( E_{z,{\rm{appro}}}^{\rm{comm}}[n] + E_{z,{\rm{appro}}}^{\rm{comp}}[n] \right) \le E_{z,{\rm{com}}}^{\max },  \forall z, \label{p4.1Ecom} \\
&\sum_{t = 1}^{2N} E_{m,{\rm{appro}}}^{\rm{prop}}[t] \le E_{m,{\rm{prop}}}^{\max },  \forall m. \label{p4.1Efly}
\end{align}
\end{subequations}

In (\hyperref[P4.1]{P4.1}), based on \eqref{eq30} and \eqref{newEcomp}, it can be observed that the computation delay and computation energy consumption are primarily influenced by the offloading decision and the allocated computation frequency. As the allocated computation frequency increases, the computation delay decreases, while the computation energy consumption increases. According to \eqref{eq31}, the communication delay is mainly determined by the offloading decision, where a higher communication rate between two nodes leads to a lower communication delay. Correspondingly, as shown in \eqref{newEcomm}, the communication energy consumption is proportional to the communication delay. Furthermore, based on \eqref{newEprop}, given a specific UAV trajectory, the UAV flight energy consumption is proportional to the flight duration. Therefore, during the optimization process, if the weight factor for delay or flight energy consumption in the system cost increases, the system tends to increase the allocated computation frequency and offload dependent sub-tasks to nearby nodes whenever possible. This reduces computation and communication delay as well as UAV flight energy consumption, while simultaneously lowering communication energy consumption and increasing computation energy consumption. Conversely, when the weight factor for computation and communication energy consumption increases, the system prioritizes lowering the computation frequency, leading to a reduction in computation energy consumption but an increase in computation delay.

\textit{3) UAV trajectories optimization:} We define $\Theta'_{3} = \{ \Theta_3,\ \{ {\tau _{2,z,z'}}[n] \}_{n,z,z'},\{ {\tau _2}[n] \}_{n} \}$. With the given ${\Theta _1}$ and ${\Theta _2}$, the subproblem of optimizing ${\Theta'_{3}}$ can be written as

\begin{subequations}
\label{P5}
\begin{align}
&{\rm{(P5):   }}\mathop {\min }\limits_{\{ \Theta'_{3}\}} {\Omega(\Theta'_{3})} , \nonumber \\
&{\rm{s.t.}\ }{\eqref{eq2},\ \eqref{eq3},\ \eqref{eq24a},\ \eqref{eq31},\ \eqref{eq33},\ \eqref{eq35c}, \eqref{eq35b}\ }, \nonumber \\
&\sum\limits_{n = 1}^N {\left( {{\tau _1}[n]P_m^{{\rm{prop}}}[2n - 1] + {\tau _2}[n]P_m^{{\rm{prop}}}[2n]} \right)}  \le E_{m,{\rm{prop}}}^{{\rm{max}}},  \forall  m. \label{eq41a}
\end{align}
\end{subequations}

In (\hyperref[P5]{P5}), both the objective function and the constraints \eqref{eq3}, \eqref{eq31}, and \eqref{eq41a} are non-convex. Specifically, in \eqref{eq3}, based on the local points ${\bf{q}}_m^j[t]$ and ${\bf{q}}_m'^j[t]$ in the $j$-th iteration, the lower bound of $||{{\bf{q}}_m}[t] - {{\bf{q}}_{m'}}[t]|{|^2}$ can be obtained by performing a first-order Taylor expansion, then we can reformulate \eqref{eq3} as 
{\begin{align}
& d_{\min }^2  \le ||{\bf{q}}_m^j[n] - {\bf{q}}_{m'}^j[n]|{|^2} \nonumber + 2{( {{\bf{q}}_m^j[n] - {\bf{q}}_{m'}^j[n]} )^ \top }\\
& \ \ \ \ \ \ \ \ \ ( {{{\bf{q}}_m}[n] + {\bf{q}}_{m'}^j[n] - {{\bf{q}}_{m'}}[n] - {\bf{q}}_m^j[n]}). \label{eq42}
\end{align}}\par

For constraint \eqref{eq31}, we introduce  ${\omega _{zz'}}[n] = \sum\limits_u {\sum\limits_k {\sum\limits_{k'} {x_{u,z}^kx_{u,z'}^{k'}o_u^{kk'}} } } $, where $\forall z,z',v_u^k \in v[n],v_u^{k'} \in {\rm{suc}}(v_u^k)$. Furthermore, we define ${\alpha _z} = \frac{{{\beta _0}{p_z}}}{{\bar B{N_0}}}$, where $\forall z \in {{\cal Z}}$. The squared distance between any two devices is given by ${d_{zz'}}[t]$.  Specifically, when $z \in {\cal M}, z' \in \{M+1,...,M+U\}$, ${d_{zz'}}[t]={d^2_{mu}}[t]$, and when $z,z' \in {\cal M}, z' \ne z$, ${d_{zz'}}[t]={d^2_{mm'}}[t]$. Based on the above definitions, we introduce slack variables ${y_{zz'}}[n]$ and ${\pi _{zz'}}[n]$ as
{\begin{align}
&{y_{zz'}}[n] \ge {d_{zz'}}[2n],{\rm{  }}\forall n, z,z',z \ne z', \label{eq44}\\
&{\pi _{zz'}}[n] \le {\log _2}\left( {1 + \frac{{{\alpha _{zz'}}}}{{{y_{zz'}}[n]}}} \right),{\rm{  }}\forall n, z,z',z \ne z', \label{eq45}
\end{align}}where the right-hand side term in \eqref{eq45} is convex, making it amenable to SCA. Similar to \eqref{eq42}, \eqref{eq45} can be transformed as

{\setlength\abovedisplayskip{2pt}
{\setlength\belowdisplayskip{2pt}
{\begin{align}
&{\pi _{zz'}}[n]  \le {\log _2}(1 + \frac{{{\alpha _{zz'}}}}{{y_{zz'}^j[n]}}) - \frac{{{\alpha _{zz'}}({y_{zz'}}[n] - y_{zz'}^j[n])}}{{y_{zz'}^j[n]({\alpha _{zz'}} + y_{zz'}^j[n])\log (2)}},\nonumber \\
&\forall n, z, z',z \ne z'. \label{eq46}
\end{align}}}}\par

Then, constraint \eqref{eq31} can be rewritten as
{\begin{equation}
{\tau _{2,z,z'}}[n] \ge \frac{{{\omega _{zz'}}[n]}}{{\bar B{\pi _{zz'}}[n]}},{\rm{ }}\forall n, z,z',z \ne z'. \label{eq47}
\end{equation}}\par

Based on Eq. \eqref{eq47}, it can be analyzed that in order to reduce the communication delay, the UAVs will adjust their trajectories to be closer to each other based on the dependency between the sub-tasks they are carrying, thus increasing the communication rate.

For constraint \eqref{eq41a}, we define ${d_m}[t] \ge ||{{\bf q}_m}[t + 1] - {{\bf q}_m}[t]||$, let $E_{m,1}^{{\rm{fly}}}[n]$ and $E_{m,2}^{{\rm{fly}}}[n]$ respectively represent the flight power consumption during computation unit ${\tau ^{{\rm{comp}}}}[n]$ and communication unit ${\tau ^{{\rm{comm}}}}[n]$,  as indicated in \eqref{eq48} and \eqref{eq49}. \par

\newcounter{TempEqCnt3} 
\setcounter{TempEqCnt3}{\value{equation}}
\setcounter{equation}{54}
\begin{figure*}[hb]
\hrulefill
\begin{align}
E_{m,1}^{{\rm{fly}}}[n] &= {P_0}\left( {1 + \frac{{3{d_m}{{[2n - 1]}^2}}}{{U_{tip}^2{\tau _1}[n]}}} \right) + \frac{{{d_0}\beta As{d_m}{{[2n - 1]}^3}}}{{2{\tau _1}{{[n]}^2}}} + {\tau _1}[n]{P_i}{\left( {\sqrt {1 + \frac{{{v_m}{{[2n - 1]}^4}}}{{4v_0^2}}}  - \frac{{{v_m}{{[2n - 1]}^2}}}{{2v_0^2}}} \right)^{\frac{1}{2}}},  \forall n, m,\label{eq48}\\
E_{m,2}^{{\rm{fly}}}[n] &= {P_0}\left( {1 + \frac{{3{d_m}{{[2n]}^2}}}{{U_{tip}^2{\tau _2}[n]}}} \right) + \frac{{{d_0}\beta As{d_m}{{[2n]}^3}}}{{2{\tau _2}{{[n]}^2}}} + {\tau _2}[n]{P_i}{\left( {\sqrt {1 + \frac{{{v_m}{{[2n]}^4}}}{{4v_0^2}}}  - \frac{{{v_m}{{[2n]}^2}}}{{2v_0^2}}} \right)^{\frac{1}{2}}}, \forall n, m. \label{eq49}
\end{align}
\end{figure*}
\setcounter{equation}{\value{TempEqCnt3}} 

\setcounter{equation}{56}

To tackle the third part of Eq. \eqref{eq48} and Eq. \eqref{eq49}, which are non-convex, we further introduce slack variable $v_m[t] \ge \| \mathbf{v}_m[t] \|$. Note that for a computation unit $t = 2n - 1$, ${v_m}[2n - 1] \ge \left( {||{{\bf q}_m}[2n] - {{\bf q}_m}[2n - 1]||} \right)/{\tau _1}[n]$ is convex since ${\tau _1}[n]$ is fixed. However, for a communication unit $t=2n$, ${v_m}[2n] \ge \left( {||{{\bf q}_m}[2n + 1] - {{\bf q}_m}[2n]||} \right)/{\tau _2}[n]$ is non-convex. Define ${{\widehat \tau } _2}[n] \ge 1/{\tau _2}[n]$, based on the given points $\widehat \tau_2^j[n]$ and $d_z^j[2n]$, similar to \eqref{eq37}, through Taylor first-order expansion, we can get
{\setlength\abovedisplayskip{2pt}
{\setlength\belowdisplayskip{2pt}
{\begin{align}
{v_m}[2n] &\ge \frac{{{{\left( {{d_m}[2n] + {{{\widehat \tau } }_2}[n]} \right)}^2}}}{2} - \frac{{2{d_m}[2n]d_z^j[2n] - {{\left( {d_m^j[2n]} \right)}^2}}}{2} \nonumber \\
 & - \frac{{2{{{\widehat \tau } }_2}[n]{\widehat \tau } _2^j[n] - {{\left( {{\widehat \tau } _2^j[n]} \right)}^2}}}{2}, \forall n,m. \label{eq50}
\end{align}}}}\par

In the same vein as \eqref{eq40}, we introduce slack variable ${u_m}{[t]^2} \ge \sqrt {1 + \frac{{{v_m}{{[t]}^4}}}{{4v_0^4}}}  - \frac{{{v_m}{{[t]}^2}}}{{2v_0^2}}$, which can be simplified to $\frac{1}{{{u_m}{{[t]}^2}}} \le {u_m}{[t]^2} + \frac{{{v_m}{{[t]}^2}}}{{v_0^2}}$, by applying Taylor first-order expansion, we can obtain
{\begin{align}
\frac{1}{{{u_m}{{[t]}^2}}} \le & - u_m^j{[t]^2} + 2{u_m}[t]u_m^j[t]\nonumber \\
& - \frac{{v_m^j{{[t]}^2}}}{{{v_0}}} + \frac{{2{v_m}[t]v_m^j[t]}}{{{v_0}}},\forall t,m. \label{eq51}
\end{align}} \par

Based on the above discussion, Eq. \eqref{eq48}, Eq. \eqref{eq49} can be rewritten as
{\begin{align}
&E_{m,1}^{{\rm{fly}}}[n] = {P_0}\left( {1 + \frac{{3{d_m}{{[2n - 1]}^2}}}{{U_{tip}^2{\tau _1}[n]}}} \right) + \frac{{{d_0}\beta As{d_m}{{[2n - 1]}^3}}}{{2{\tau _1}{{[n]}^2}}}\nonumber \nonumber\\
 &\ \ \ \ \ \ \ \  + {\tau _1}[n]{P_i}{u_m}[2n - 1],\ \ \ \  \forall n, m, \label{eq52}\\
&E_{m,2}^{{\rm{fly}}}[n] = {P_0}\left( {1 + \frac{{3{d_m}{{[2n]}^2}}}{{U_{tip}^2{\tau _2}[n]}}} \right) + \frac{{{d_0}\beta As{d_m}{{[2n]}^3}}}{{2{\tau _2}{{[n]}^2}}}\nonumber \\
 &\ \ \ \ \ \ \ \  + {\tau _2}[n]{P_i}{u_m}[2n], \ \ \ \  \forall n, m. \label{eq53} 
\end{align}}\par

Furthermore, similar to \eqref{eq36}, we can transform the non-convex component  ${\tau _2}[n]{u_m}[2n]$ in \eqref{eq54} into
{\begin{align}
& ({\tau _2}[n]{u_m}[2n])_{\rm appro}  = \frac{{{{\left( {{\tau _2}[n] + {u_m}[2n]} \right)}^2}}}{2} - \frac{{2{\tau _2}[n]\tau _2^j[n] }}{2}\nonumber \\
 & \quad \quad \quad - \frac{{2{u_m}[2n]u_m^j[2n]}}{2} + \frac{{{( {\tau _2^j[n]} )}^2}+{{( {u_m^j[2n]} )}^2}}{2}. \label{eq54}
\end{align}}Then, we can get

{\begin{align}
&E_{m,2,{\rm appro}}^{{\rm{fly}}}[n] = {P_0}\left( {1 + \frac{{3{d_m}{{[2n]}^2}}}{{U_{tip}^2{\tau _2}[n]}}} \right) + \frac{{{d_0}\beta As{d_m}{{[2n]}^3}}}{{2{\tau _2}{{[n]}^2}}}\nonumber \\
 &\ \ \ \ \ \ \ \  + {P_i}({\tau _2}[n]{u_m}[2n])_{\rm appro}, \ \ \ \  \forall n, m. \label{eq53} 
\end{align}}\par

By incorporating the above derivations into \eqref{eq21}, the UAV flight energy consumption can be reformulated as

\begin{align}
E_m^{{\rm{prop}}}[n] = E_{m,1}^{{\rm{fly}}}[n] + E_{m,2,{\rm{appro}}}^{{\rm{fly}}}[n], \forall n, m.
\label{newEq21}
\end{align}Accordingly, the energy constraint in Eq.~\eqref{eq41a} can be rewritten as

\begin{align}
\sum_{n} (E_{m,1}^{{\rm{fly}}}[n] + E_{m,2,{\rm{appro}}}^{{\rm{fly}}}[n]) \leq E_{z, \rm prop}^{\rm max}, \forall m.
\end{align} \par

Based on the above transformations and approximations, the original non-convex problem (\hyperref[P5]{P5}) is now equivalently reformulated as a convex optimization problem, which can be efficiently solved using standard convex solvers.

In (\hyperref[P5]{P5}), based on \eqref{eq47}, \eqref{eq52}, and \eqref{eq53}, it can be observed that when the offloading decision and computation frequency allocation are fixed, the optimization of UAV trajectories primarily affects communication delay, communication energy consumption, and UAV flight energy consumption. Specifically, increasing the weight factor for system delay or communication energy consumption drives UAVs to plan trajectories that reduce inter-node distances. This proximity enhances communication rates, thereby reducing both communication delay and energy consumption. Furthermore, when the weight factor for UAV flight energy consumption rises, trajectories are adjusted to lower flight speeds while maintaining minimal communication delay, aiming to reduce overall flight energy consumption. However, in typical computing scenarios where computation delay often exceeds communication delay, UAV flight energy consumption remains subject to fluctuations in computation delay, even when prioritizing flight energy efficiency.

\subsection{PDD-SCA Algorithm Design and Analysis}

In the outer loop of the PDD-SCA algorithm, by solving (\hyperref[P3]{P3}), the AL multipliers are updated as follows
{\begin{align}
    \lambda _{u,k,z}^{1,l + 1} &= \lambda _{u,k,z}^{1,l} + \frac{1}{{{\varrho ^n}}}\left( {x_{u,z}^k\left( {\tilde x_{u,z}^k - 1} \right)} \right), \forall u, k, z,  \label{eq55}
\end{align}}
{\begin{align}
        \lambda _{u,k,z}^{2,l + 1} &= \lambda _{u,k,z}^{2,l} + \frac{1}{{{\varrho ^l}}}\left( {x_{u,z}^k - \tilde x_{u,z}^k} \right),\forall u, k, z,  \label{eq56}
\end{align}}
where $l$ denotes the number of iterations in the previous loop.\par

According to~\cite{ref38} and~\cite{ref40}, we define $\Delta  = \{ {\bf X},{\bf \tilde X}\} $ and introduce an indicator function $h(\Delta )$. This function measures the extent of violation for the constraints  \eqref{eq25}, \eqref{eq26} in sub-task offloading variables and signals the termination condition of the algorithm, as given by
{\begin{equation}
h({\Delta^l}) = \max \left\{ {{{\left\| {x_{u,z}^k(\tilde x_{u,z}^k - 1)} \right\|}^2},{{\left\| {x_{u,z}^k - \tilde x_{u,z}^k} \right\|}^2}} \right\}. \label{eq57}
\end{equation}}\par

The proposed PDD-SCA algorithm, outlined in Algorithm 1, consists of an outer loop and an inner loop. The inner loop procedure is detailed in Algorithm 2. Specifically, we first adopt the PDD framework to relax the original nonconvex problem (\hyperref[P1]{P1}) involving binary variables by introducing auxiliary variables, penalty parameter, and AL multipliers. In the outer loop, the penalty parameter and AL multipliers are updated iteratively. Given fixed values of these parameters, the inner loop sequentially solves three subproblems: (\hyperref[P3]{P3}) is convex, while the (\hyperref[P4]{P4}) and (\hyperref[P5]{P5}) are nonconvex. For  (\hyperref[P4]{P4}) and (\hyperref[P5]{P5}), we construct surrogate upper-bound functions by applying a first-order Taylor expansion to the nonconvex terms, thereby transforming them into approximately convex subproblems. This Block-SCA procedure guarantees a monotonic decrease of the objective value. As the iterations proceed, the inner loop generates stationary points of the relaxed problem under fixed penalty parameter and AL multipliers~\cite{SCA}. Furthermore, as the outer loop iteratively refines the penalty and AL multipliers, the penalty term becomes sufficiently tight to enforce constraint satisfaction, eventually making the relaxed problem equivalent to the original one. In this way, the algorithm converges to a KKT point of the original problem~\cite{ref38}. In addition, since the objective is defined as a weighted sum of delay and energy consumption, the obtained solution corresponds to a Pareto-efficient point in the context of multi-objective optimization~\cite{weightedSumMOO}.

For the proposed PDD-SCA algorithm, the outer loop updates the AL multipliers and penalty parameter, while the inner loop alternately optimizes sub-task offloading, computational resource allocation, and UAV trajectories. The complexity of the algorithm mainly comes from the SCA algorithm involved in the inner loop. Specifically, let $\kappa $ denote the average number of sub-tasks of each task in each time slot. Let ${I_1}$ represent the number of iterations of the inner loop. Based on~\cite{ref4,ref42}, the complexity of Algorithm \ref{alg1} is approximated as $O({I_1}{N^3}{U^3}{M^3}{\kappa ^3})$. Furthermore, let ${I_2}$ be the number of iterations of the outer loop. Therefore, the complexity of the dual-loop PDD-SCA algorithm in Algorithm \ref{alg2} is calculated as $O({I_1}{I_2}{N^3}{U^3}{M^3}{\kappa ^3})$.

\begin{algorithm}[t]
	\caption{The Proposed Dual-Loop PDD-SCA Algorithm}
	\label{alg2}
	\begin{algorithmic}
		\STATE Initialize original variables ${\Theta _1},{\Theta _2},{\Theta _3},\lambda _{u,k,z}^1,\lambda _{u,k,z}^2,\varrho $, set the tolerance of accuracy $\varepsilon $ and $\varsigma $ for the inner and outer loop, set $0 < c < 1$, ${\eta ^0} > 0$, give the maximum number of iteration ${I_{{\rm{2,max}}}}$, and the current iteration number $l = 0$.
		\REPEAT
		\STATE 1: Update ${\Theta _1},{\Theta _2},{\Theta _3}$ via \textbf{Algorithm \ref{alg1}};
        \STATE     2: \textbf{if} ${\left\| {h({\Delta ^l})} \right\|_\infty } \le {\eta ^l}$ \textbf{then}
        \STATE \hspace{0.8cm}          Update AL multipliers according to \eqref{eq55}, \eqref{eq56},
        \STATE \hspace{0.8cm}  ${\varrho ^{l + 1}} = {\varrho ^l}$ .
         \STATE  \hspace{0.3cm}  \textbf{else}
        \STATE \hspace{0.8cm} 		update ${\varrho ^{l + 1}}$ by decreasing ${\varrho ^l}:{\varrho ^{l + 1}} = c{\varrho ^l}$.
        \STATE    \hspace{0.3cm}  \textbf{endif}
		\STATE 3: Update ${\eta ^{l + 1}} = 0.7{\left\| {h({\Delta ^l})} \right\|_\infty }$;
		\STATE 4: Update $l \leftarrow l + 1$;
		\UNTIL reach the accuracy ${\left\| {h({\Delta ^l})} \right\|_\infty } < \varsigma $ or exceed the value of $l \ge {I_{{\rm{2,max}}}}$.
	\end{algorithmic}  
\end{algorithm}

\section{Numeric Results}
In this section, we present the simulation results to verify the performance of the proposed PDD-SCA algorithm.

\textit{1) System settings:} We consider a rectangular area with dimensions of  $200\ {\rm m} \times  200\ {\rm m}$, where the TDs are randomly distributed within three circular regions, each with a radius of $40\ {\rm m}$. The UAVs are deployed at the midpoint between the center lines of every two adjacent TDs' areas.\par

\begin{algorithm}[t]
	\caption{Inner-Loop Update of PDD-SCA Algorithm}
	\label{alg1}
	\begin{algorithmic}
		\STATE Initialize original variables ${\Theta _1},{\Theta _2},{\Theta _3},\lambda _{u,k,z}^1,\lambda _{u,k,z}^2,\varrho $, set the tolerance of accuracy $\varepsilon $, give the maximum number of iteration ${I_{{\rm{1,max}}}}$, and the current iteration number $j=0$.
		\REPEAT
		\STATE 1: Update ${\Theta _1}$ based on $\eqref{eq29}$;
		\STATE 2: Update ${\Theta _2}$ based on solving (\hyperref[P4.1]{P4.1});
		\STATE 3: Update ${\Theta _3}$ based on solving (\hyperref[P5.1]{P5.1});
		\STATE 4: Update $j \leftarrow j + 1$
		\UNTIL the difference between consecutive values of the objective function is under $\varepsilon $ or $j \ge {I_{{\rm{1,max}}}}$.
	\end{algorithmic}  
\end{algorithm}

For computational tasks, we consider face recognition applications that can be used for airport security and border defense  (which can be subdivided into multiple sub-tasks, e.g., coarsely into image processing, feature extraction, feature fusion, and predictive classification, where the output data of the preceding sub-tasks needs to be transmitted to the succeeding sub-tasks)~\cite{computationtask1}. Referring to the data in~\cite{computationtask1,computationtask2}, the amount of input data for each task is set between $[1.5,\ 3]\ {\rm Mbits}$~\cite{computationtask1}.  The number of time slots is set to $N=20$, and the sub-task topology is constructed layer by layer as described in~\cite{ref19}, with the number of layers equaling the number of time slots. The number of sub-tasks in each layer is randomly selected from a uniform distribution $[1,\ 3]$. The workload of each sub-task is randomly determined from a uniform distribution $[0.8,\ 1]\ {\rm Mbits}$. Based on these settings, the total computational volume of each task is approximately $32.4\ {\rm Mbits}$, aligning closely with the setting in~\cite{computationtask2}. For dependencies between sub-tasks, without loss of generality, dependency modeling is performed with random probabilities, and the number of succeeding dependent sub-tasks of each sub-task $v_u^k$ is randomly selected from the range [1, ${\rm max}(1,\ 0.02 \times (K-k))$], following a uniform distribution. The transmission data volume between any two sub-tasks with dependency is generated from a uniform distribution ranging between [0.1, 0.2] ${\rm Mbits}$. Furthermore, considering that the deadline for each task is related to its computational volume and computing frequency, we set the deadline to  $( {{\sum\limits_k {c_u^k{\vartheta _z}} } } )/(1.5 \times {10^9})$, ensuring a minimum allocated computation frequency of $1.5\ {\rm GHz}$ for each task.\par

\begin{table}[!t]
\renewcommand{\arraystretch}{1.4}
\caption{Simulation Parameters\label{tab:table2}}
\label{table2}
\centering
\begin{tabular}{|p{5.5cm}|c|}
\hline
\textbf{Symbolic meaning }& \textbf{Symbol and Value}\\ \hline
Altitude of UAVs & $H=50\ \rm m$\\ \hline
Amount of UAVs & $M=3$\\ \hline
Amount of TDs & $U=4$\\ \hline
Amount of time slots & $N=20$  \\ \hline
\multicolumn{1}{|m{5.5cm}|}{Maximum length of the communication unit} & $\tau _{\max }^{{\rm{comm}}}=0.5\ \rm s$  \\ \hline
Maximum UAV speed & ${V_{\max }}=35\ \rm m/s$\\ \hline
Minimum distance between UAVs & ${d_{\min }}=10\rm m$ \\ \hline
System bandwidth & $B=120\  \rm MHz$\\ \hline
Reference channel power & ${\beta _0}=-50\  \rm dB$\\ \hline
Transmit power of TD & $P_u^{\max }=23\ \rm dBm$\\ \hline
Transmit power of UAV & $P_m^{\max }=35\ \rm dBm$\\ \hline
Noise power spectrum density & $N_0=-130\  \rm dBm$\\ \hline
Maximum computation frequency of TD & $F_u^{\max }=0.5\  \rm GHz$\\ \hline
Maximum computation frequency of UAV & $F_m^{\max }=10\  \rm GHz$\\ \hline
\multicolumn{1}{|m{5.5cm}|}{Required CPU cycles per bit computation at TD/UAV} & ${\vartheta _z}=10^3\  \rm cycles/bit$ \\ \hline
CPU capacitance coefficient of TD/UAV & ${\gamma _z}=10^{-27}$\\ \hline
\multicolumn{1}{|m{5.5cm}|}{Weight factor of task completion delay} & $w^{{\rm{tim}}}=1$\\ \hline
\multicolumn{1}{|m{5.5cm}|}{Weight factor of TD/UAV communication and computation energy consumption} & $w_z^{{\rm{com}}}=0.09/0.01$\\ \hline
\multicolumn{1}{|m{5.5cm}|}{Weight factor of UAV flight energy consumption} & $w_m^{{\rm{fly}}}=10^{-4}$\\ \hline
\multicolumn{1}{|m{5.5cm}|}{Maximum iteration number of inner loops} &{$I_{\rm 1,max}=20$}\\ \hline
\multicolumn{1}{|m{5.5cm}|}{Maximum iteration number of outer loops} & {$I_{\rm 2,max}=50$}\\ \hline
\multicolumn{1}{|m{5.5cm}|}{Inner loop accuracy tolerance} & {$\varepsilon =10^{-3}$}\\ \hline
\multicolumn{1}{|m{5.5cm}|}{Outer loop accuracy tolerance} & {$\varsigma=10^{-10}$}\\ \hline
\multicolumn{1}{|m{5.5cm}|}{Initial value of constraint slackness} & {$\eta ^0 =10^{-3}$}\\ \hline
\multicolumn{1}{|m{5.5cm}|}{Decay coefficient} & {$c =0.9$}\\ \hline
\end{tabular}
\end{table}

Considering the energy constraints, we set the maximum computational and communication energy consumption of each TD to $10\ {\rm J}$, and set the maximum computational and communication energy consumption of each UAV to vary with the amount of computational data of the tasks (the amount of communication data of the tasks has not been taken into account because the amount of communication data is small, and the communication energy is negligible compared to the computational energy consumption), as given by $( {\sum\limits_u {\sum\limits_k {c_u^k} } {\vartheta _z}{\gamma _z}{{(3 \times {{10}^9})}^2}} )/M$, which implies that the maximum CPU frequency assigned to each task is $3\ {\rm GHz}$. For the limitation of UAV flight energy consumption, assuming that the duration of each time slot is 3 seconds and the flight speed of the UAV is 20 m/s, the maximum flight energy consumption of the UAV can be determined based on Eq. \eqref{eq19} and Eq. \eqref{eq20}.\par

Based on the settings of other system parameters and the delay and energy models in Eq. \eqref{eq16}, Eq. \eqref{eq17}, and Eq. \eqref{eq19}, we estimate the typical magnitudes of key performance metrics. Specifically, the task deadline is generally on the order of tens of seconds. The computation and communication energy consumption of each TD is roughly a few joules. For UAVs, the corresponding energy consumption is on the order of hundreds of joules, and the UAV propulsion energy is significantly higher, typically exceeding ten thousand joules, due to continuous propulsion demands. Based on these estimations, we configure the weight factors in the system cost function to ensure numerical comparability and a balanced optimization objective. Specifically, the weight factor for the task completion delay is set to $w^{\rm tim}=1$, the weight factor for the computation and communication energy consumption of TDs is set to $w_z^{\rm com} = 0.09, \forall z \in \{M+1,...,M+U\}$, that for UAVs is set to $w_z^{\rm com} = 0.01, \forall z \in \{1,...,M\}$, and the weight factor for UAV flight energy is set to $w_m^{\rm fly} = 10^{-4}$. These values are selected to normalize the scale differences among delay, device energy consumption, and flight energy consumption, ensuring that all components contribute comparably to the objective function. \par

We set the parameters of this paper with reference to the parameter settings of similar scenarios in~\cite{computationtask1, ref40, ref19, ref32}. Table \ref{table2} summarizes the main system setting parameters for the numerical simulations.\par

\textit{2) Benchmark Algorithms: } The benchmark algorithms utilize a fixed UAV trajectory, in which the UAVs will fly in a circular path with a radius of $5 \ \rm m$. These algorithms include the multiobjective evolutionary algorithm~\cite{newDependenyUAVTask1}, the heuristic offloading algorithm, the fixed time allocation algorithm and the fixed computational resource allocation algorithm.\par

\begin{itemize}
\item \textbf{Multiobjective evolutionary algorithm}: To fit the scenario of this paper, we modified the algorithm in~\cite{newDependenyUAVTask1}. Specifically, we fixed the weights of multiple optimization objectives based on the weight parameter settings in this paper. In addition, for multi-TD scenarios, we utilize the reverse breadth-first search algorithm proposed in~\cite{newDependenyTask5} to compute the priorities of sub-tasks of multiple TDs and arrange these sub-tasks into a chain of sub-tasks based on the priorities.\par

\item \textbf{Heuristic offloading algorithm}: Each TD heuristically offloads its sub-tasks to the closest UAV for task computation. Then, the UAV distributes computational resource equally among the sub-tasks it serves at each time slot.\par

\begin{figure}
    \centering
    \includegraphics[width=0.85\linewidth]{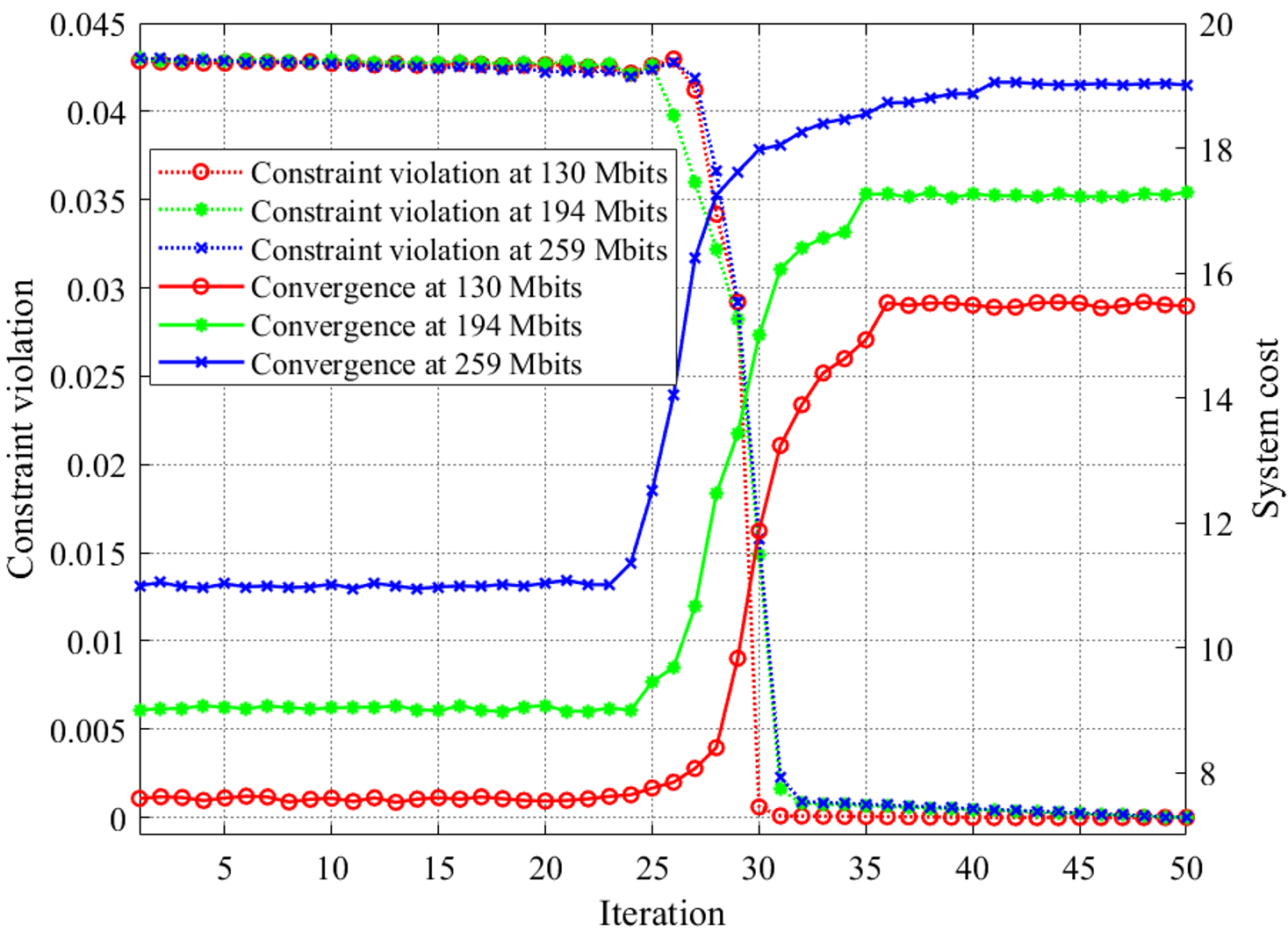}
    \caption{Convergence of the proposed algorithm.}
    \label{fig3}
\end{figure}

\begin{figure}
    \centering
    \includegraphics[width=0.75\linewidth]{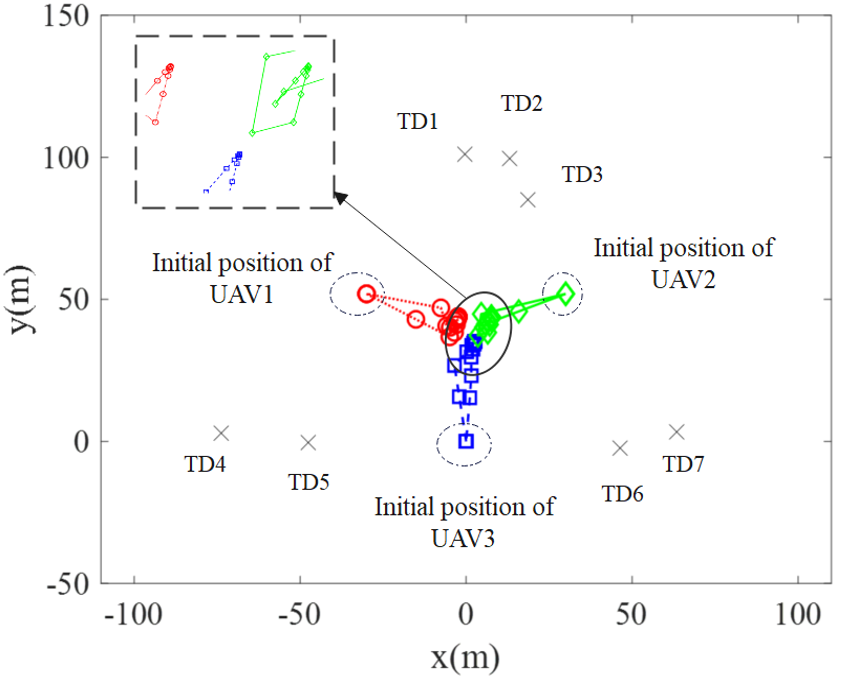}
    \caption{The optimized UAVs' trajectories.}
    \label{fig9}
\end{figure}

\item \textbf{Fixed computational resource allocation}: The parallel sub-tasks in each time slot are distributed equally to the UAVs, and each UAV distributes the CPU frequency equally to the associated sub-tasks.\par

\item \textbf{Fixed time allocation algorithm}: The parallel sub-tasks in each time slot are evenly distributed to the UAVs, and the communication unit and computational unit lengths in each time slot are fixed, based on Eq. \eqref{eq12}, the CPU frequency assigned to each sub-task can be obtained. Specifically, the communication unit and computation unit lengths of this algorithm are set according to the maximum communication unit and computation unit lengths obtained by the fixed computational resource allocation algorithm, thus ensuring that the computational resource allocated by each device in each time slot will not exceed the limit.\par

\end{itemize}

To comprehensively evaluate the proposed method, we conduct a series of simulations with different emphases. These include convergence analysis, UAV trajectory optimization, performance evaluation under varying computational workloads, communication volumes, and inter-sub-task dependency levels. Additionally, we examine the the trade-off between task completion delay and energy consumption under different weight factor settings, as well as the workload balancing performance among UAVs.

\begin{figure*}[!t]
\centering
\subfloat[System cost]{\includegraphics[width=2.08in]{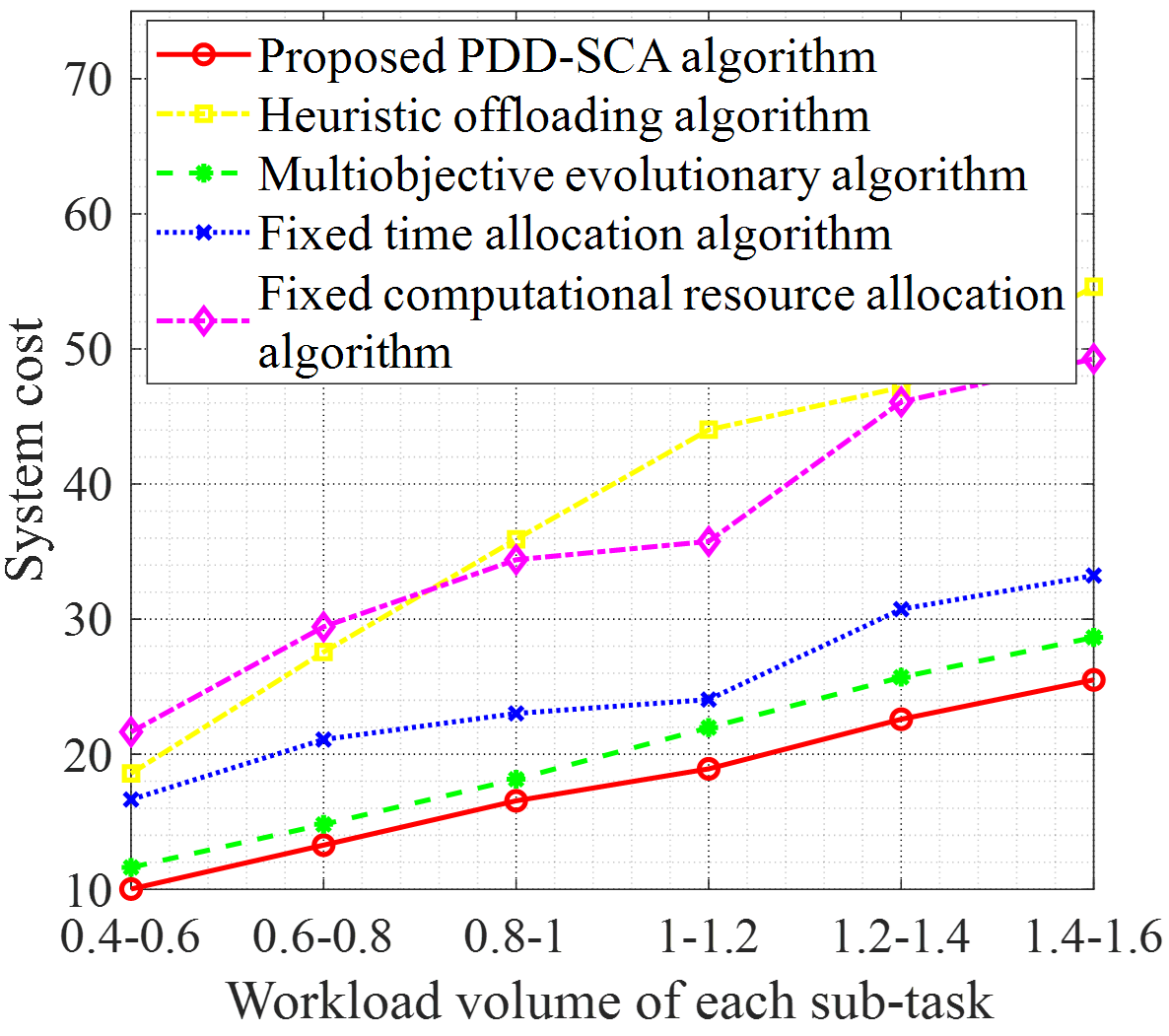}
\label{workload_sys}}%
\hfil
\subfloat[Task completion delay]{\includegraphics[width=2.2in]{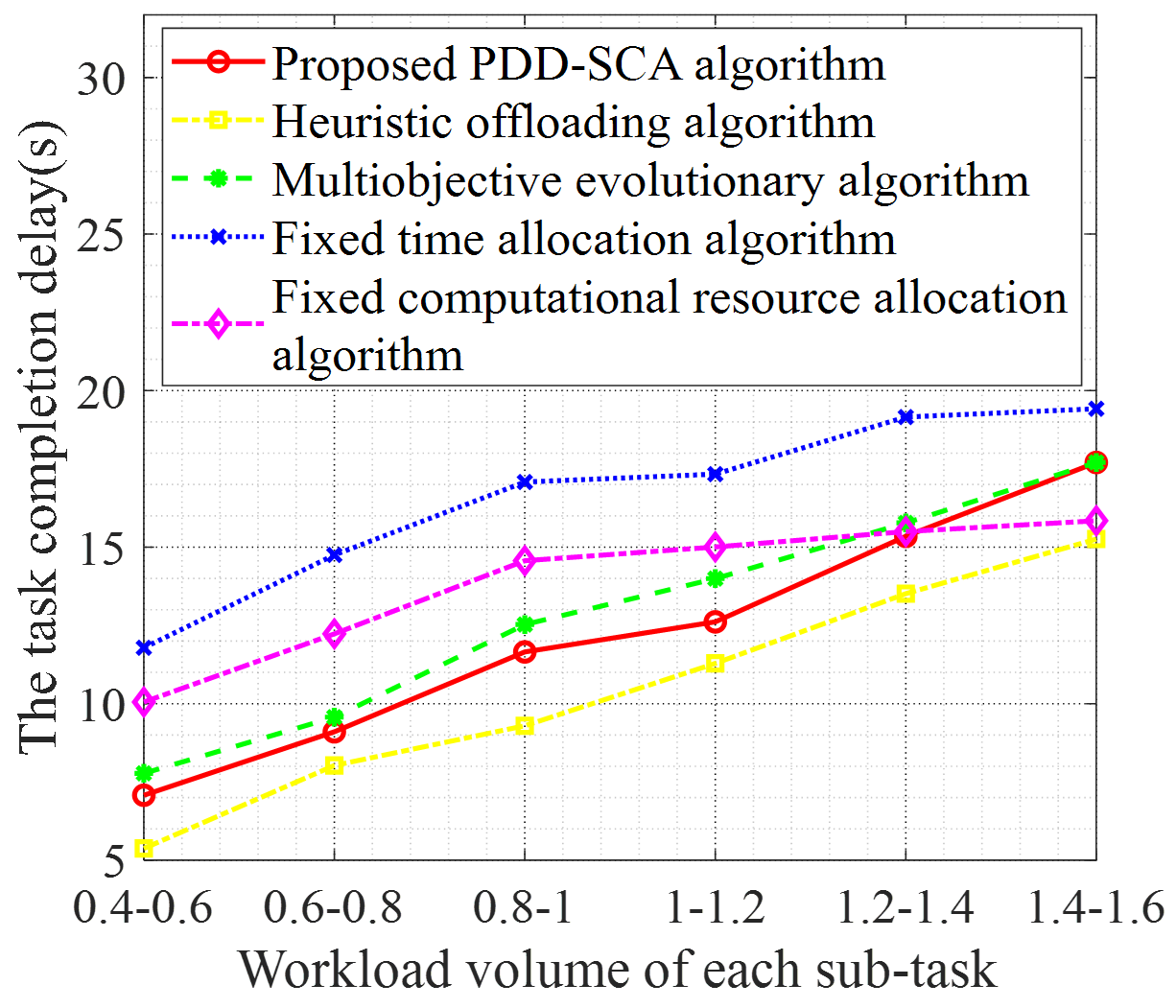}
\label{workload_delay}}%
\hfil
\subfloat[User energy consumption]{\includegraphics[width=2.3in]{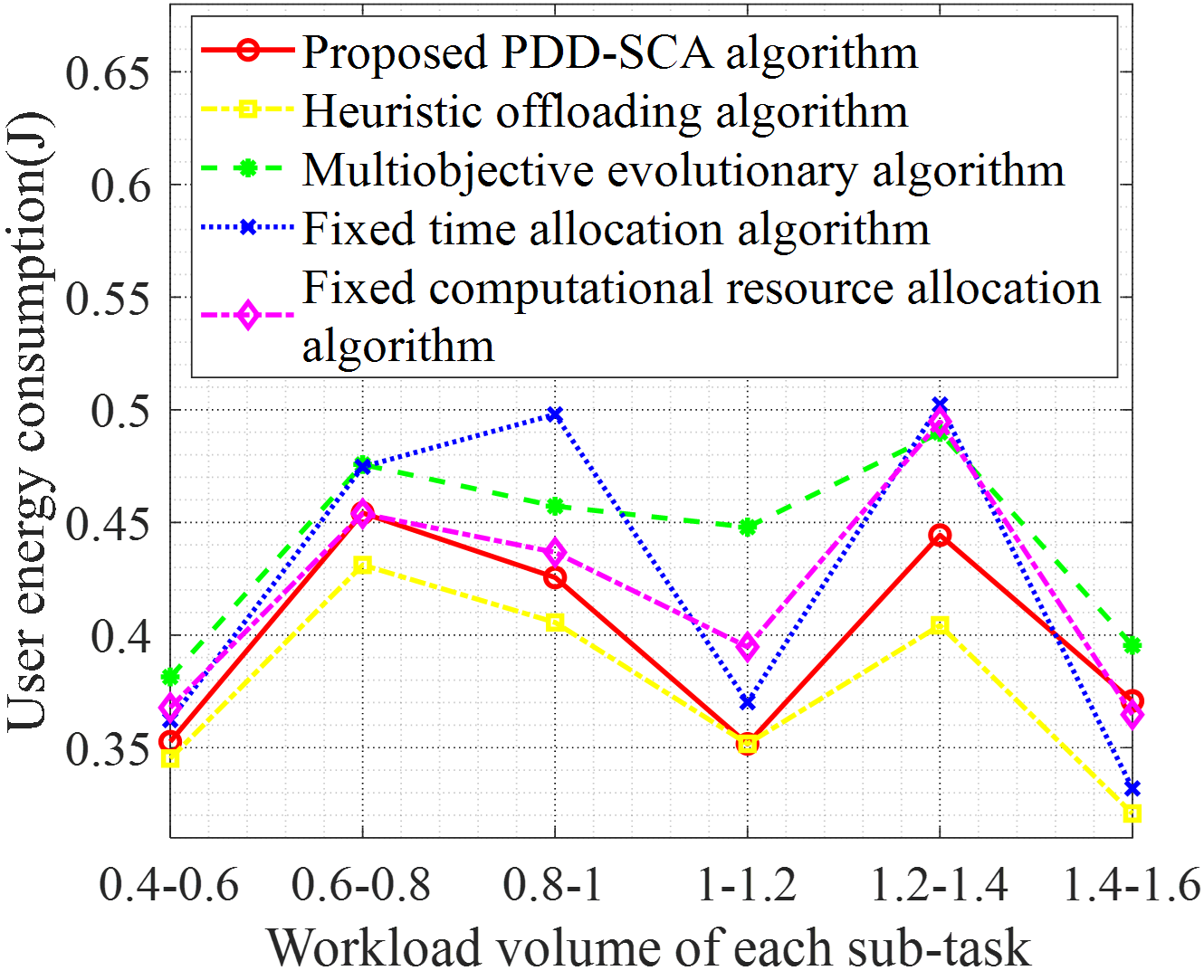}
\label{workload_eu}}%
\hfil
\subfloat[UAV computational and communication energy consumption]{\includegraphics[width=2.43in]{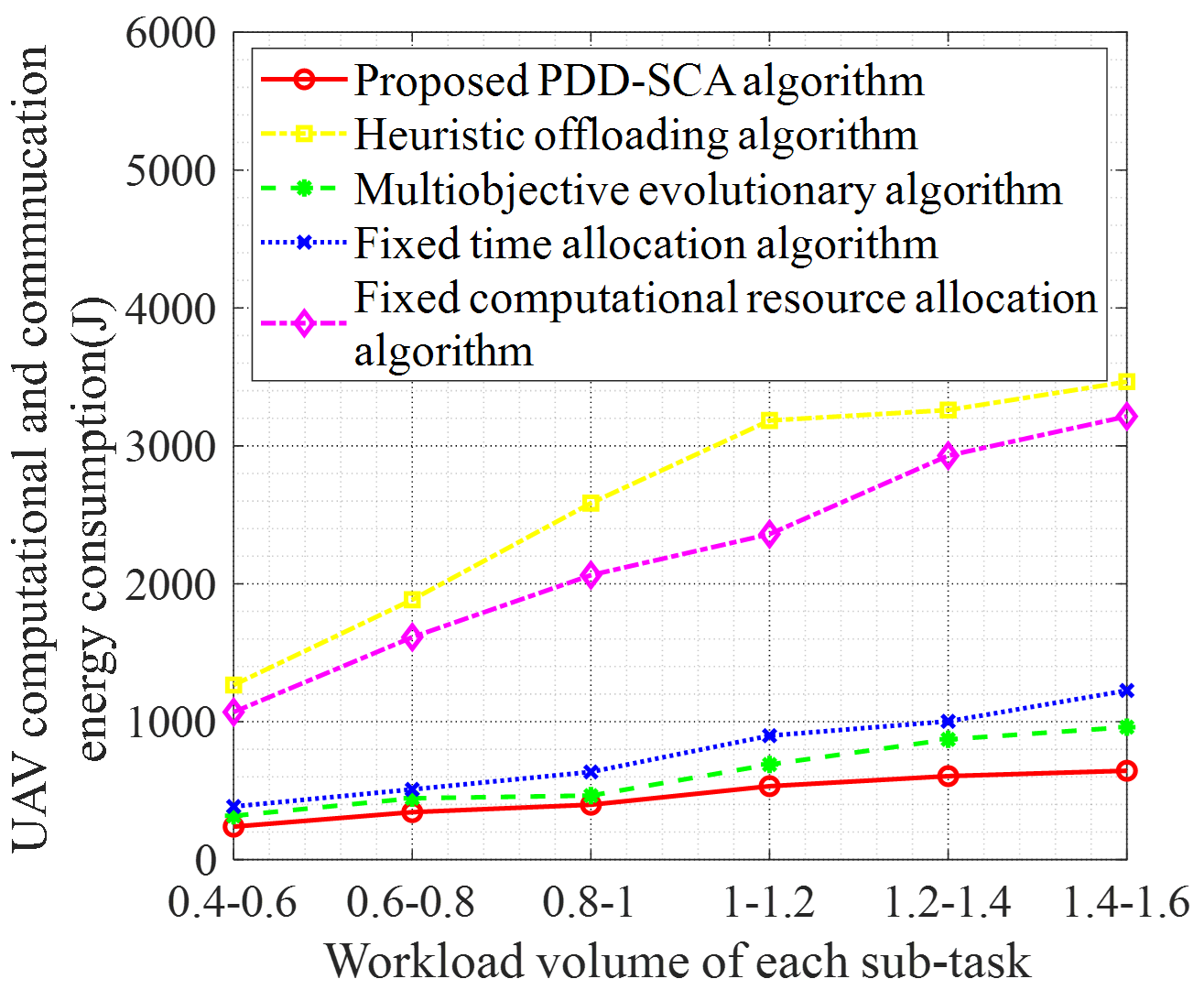}%
\label{workload_em}}
\hfil
\subfloat[UAV flight energy consumption]{\includegraphics[width=2.25in]{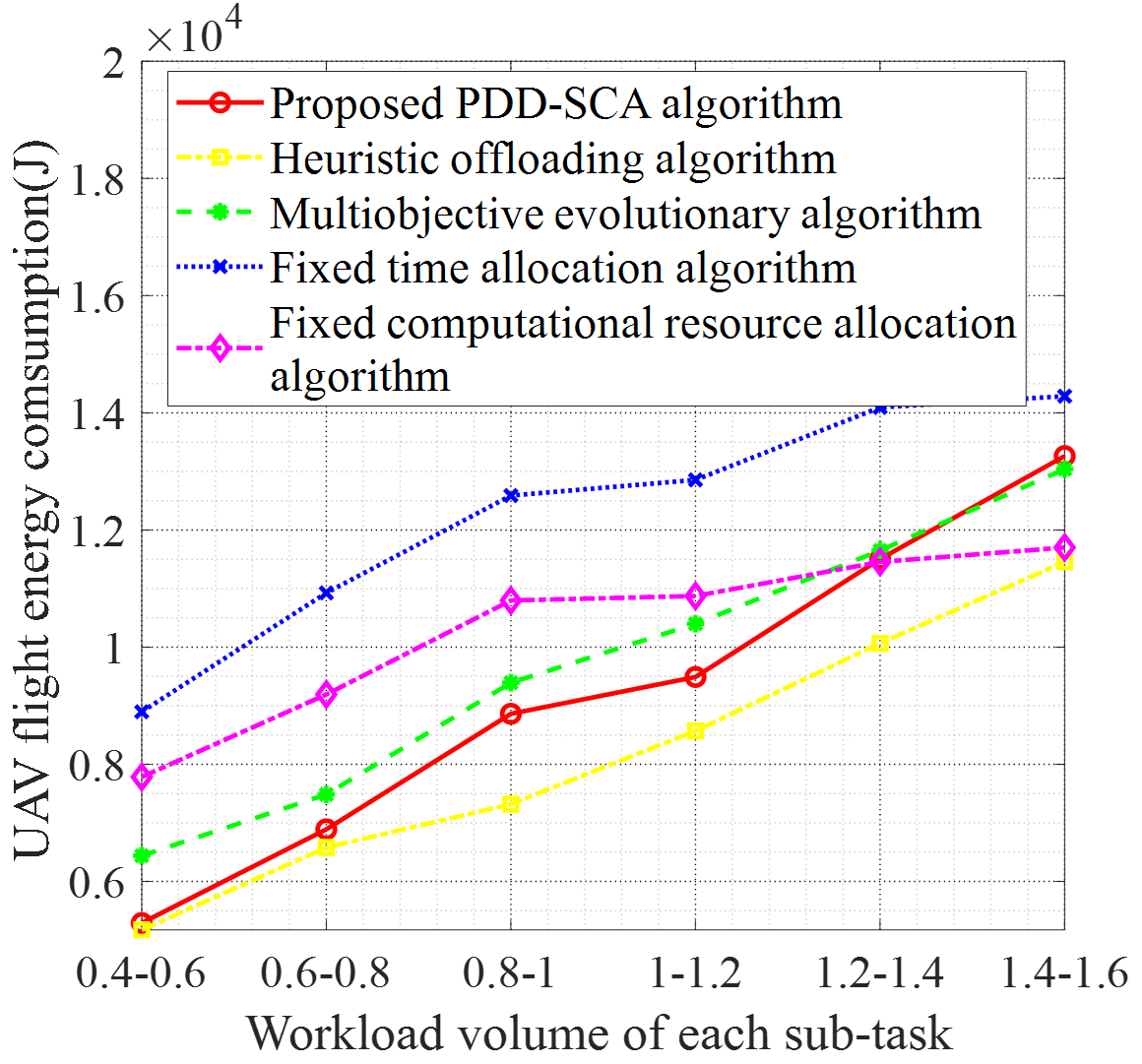}%
\label{workload_eprop}}
\caption{System performance with varying computational workloads.}
\label{fig5}
\end{figure*}

\subsection{Performance Evaluation}

Fig. \ref{fig3} illustrates the system cost convergence performance and constraint violation metrics \eqref{eq57} value variation of the PDD-SCA algorithm for three different computational load scenarios. Specifically, the number of sub-tasks per time slot involved in the three computation scenarios are set to 1-3, 2-4, and 3-5, and the corresponding average computation volumes are 130 Mbits, 194 Mbits, and 259 Mbits, respectively. As shown in the Fig. \ref{fig3}, when the number of iterations is small, the value of the constraint violation indicator is large, which means that the penalty of the binary sub-task offloading variable is small, the AL term is small, and the offloading decision variable is not approaching binary, so that the algorithm can flexibly adjust the sub-task offloading decision and the allocation of computational resource. As the number of iterations increases, the value of the constraint violation indicator gradually decreases until it approaches 0, which means that the penalty of the binary sub-task offloading variable is strengthened, resulting in a gradual increase of the AL term, and the sub-task offloading decision is approaching the binary, which makes the cost of the system increase with the increase of the delay of task completion.\par

In Fig. \ref{fig9}, we demonstrate the optimized flight trajectories of UAVs obtained using the PDD-SCA algorithm with 7 TDs. It can be observed that after UAVs receive the input data of the tasks in the first time slot, they adjust their trajectories in subsequent time slots to approach each other. This adjustment is driven by the interdependencies among UAVs' assigned sub-tasks in subsequent time slots, with the objective of mitigating both communication delay and communication energy consumption. Then, the UAVs return to the initial point at the last time slot.\par

\begin{figure*}[!t]
\centering
\subfloat[System cost]{\includegraphics[width=2.2in]{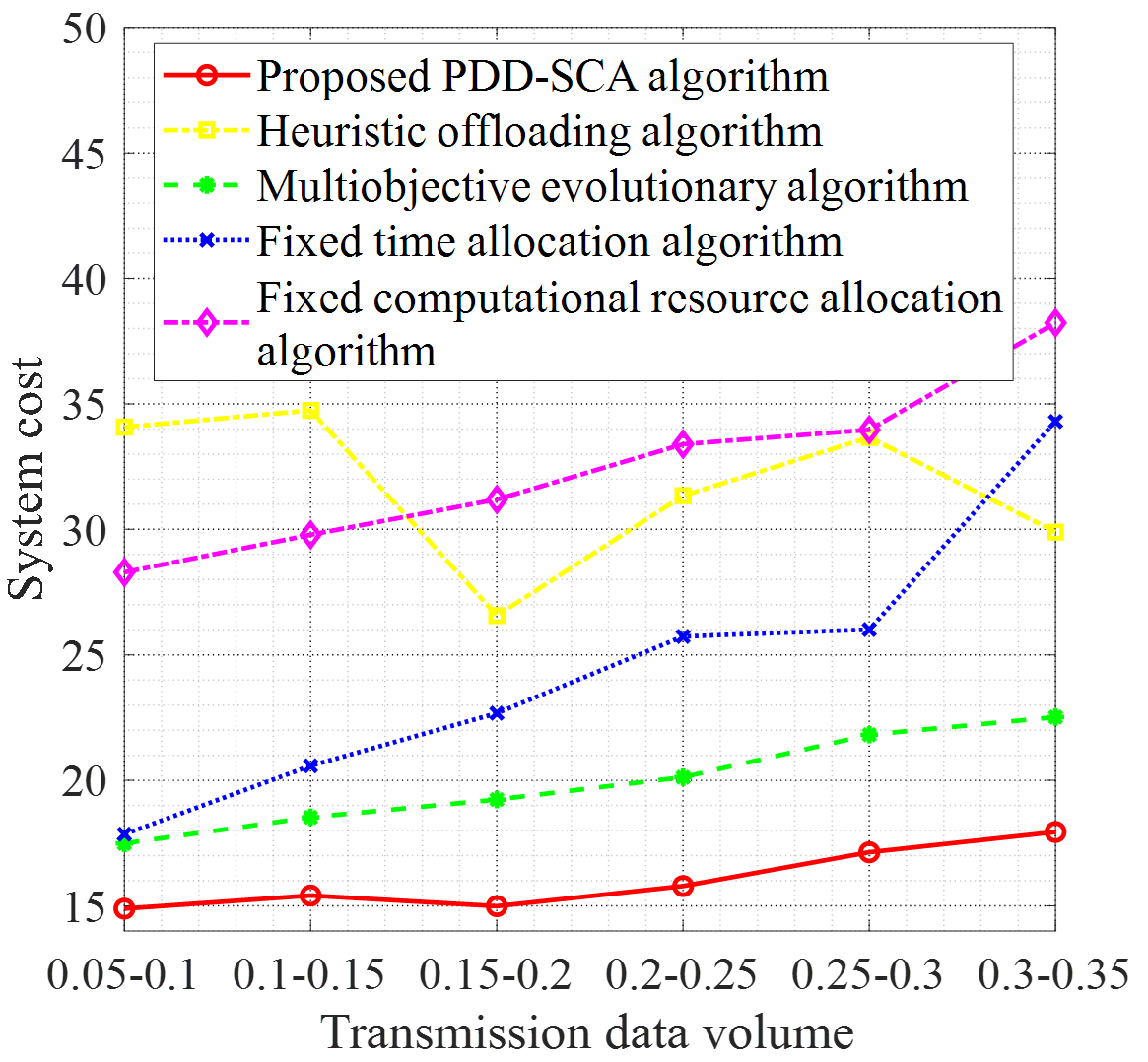}
\label{transload_sys}}%
\hfil
\subfloat[Task completion delay]{\includegraphics[width=2.36in]{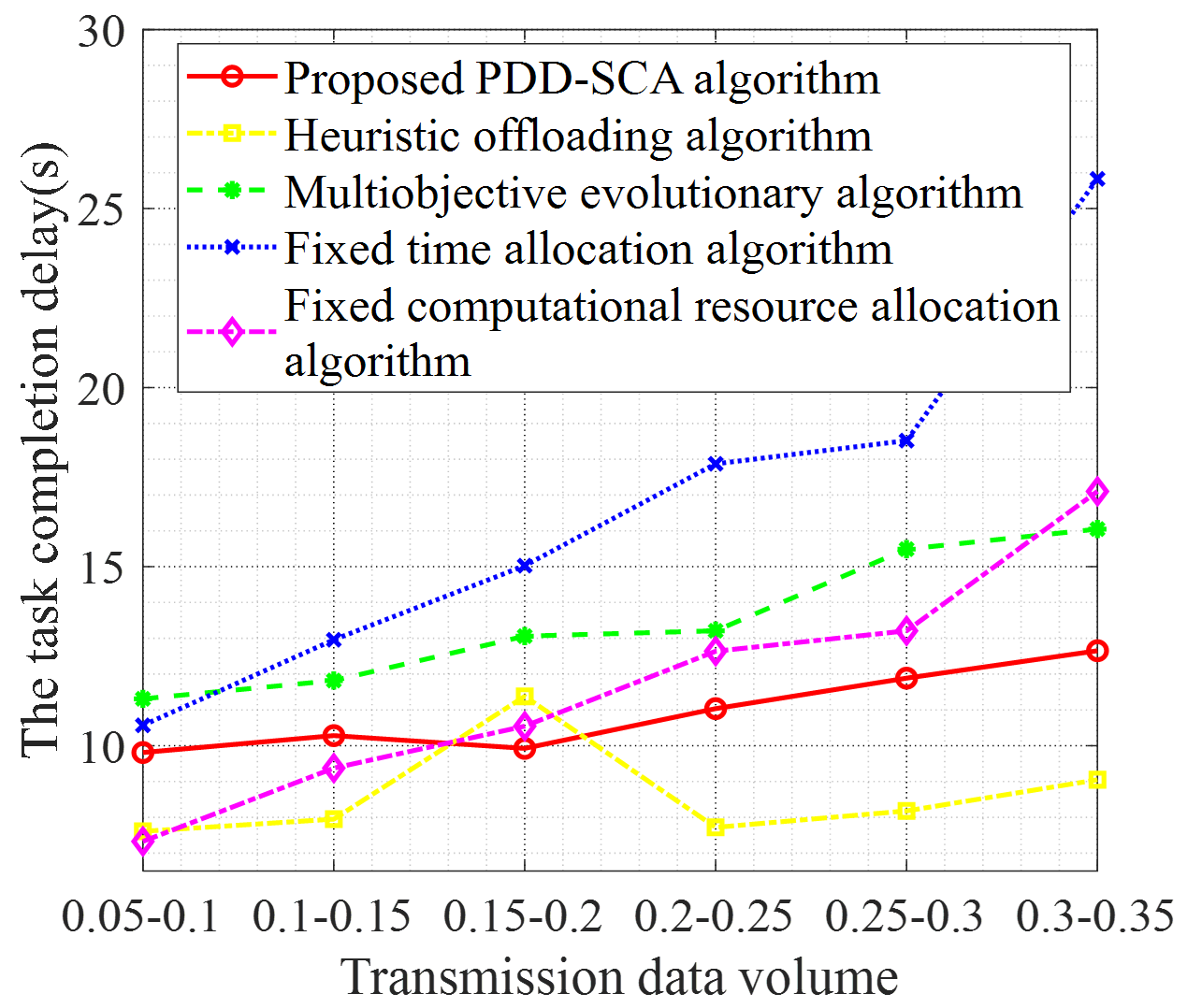}
\label{transload_delay}}%
\hfil
\subfloat[User energy consumption]{\includegraphics[width=2.36in]{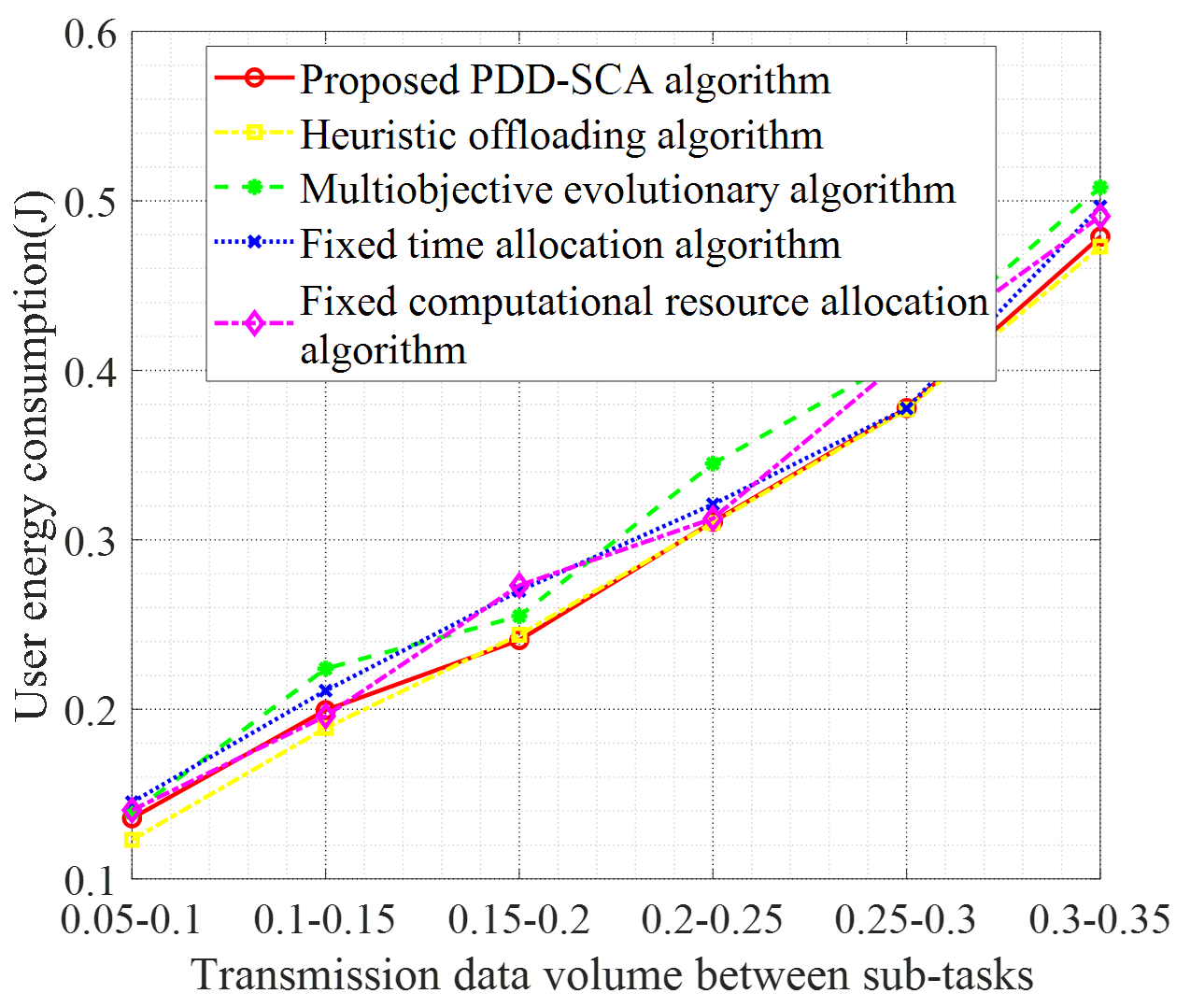}
\label{transload_eu}}%
\caption{System performance with varying inter-sub-task communication data volumes.}
\label{fig6}
\end{figure*}

\begin{figure*}[!t]
\centering
\subfloat[System cost]{\includegraphics[width=2.3in]{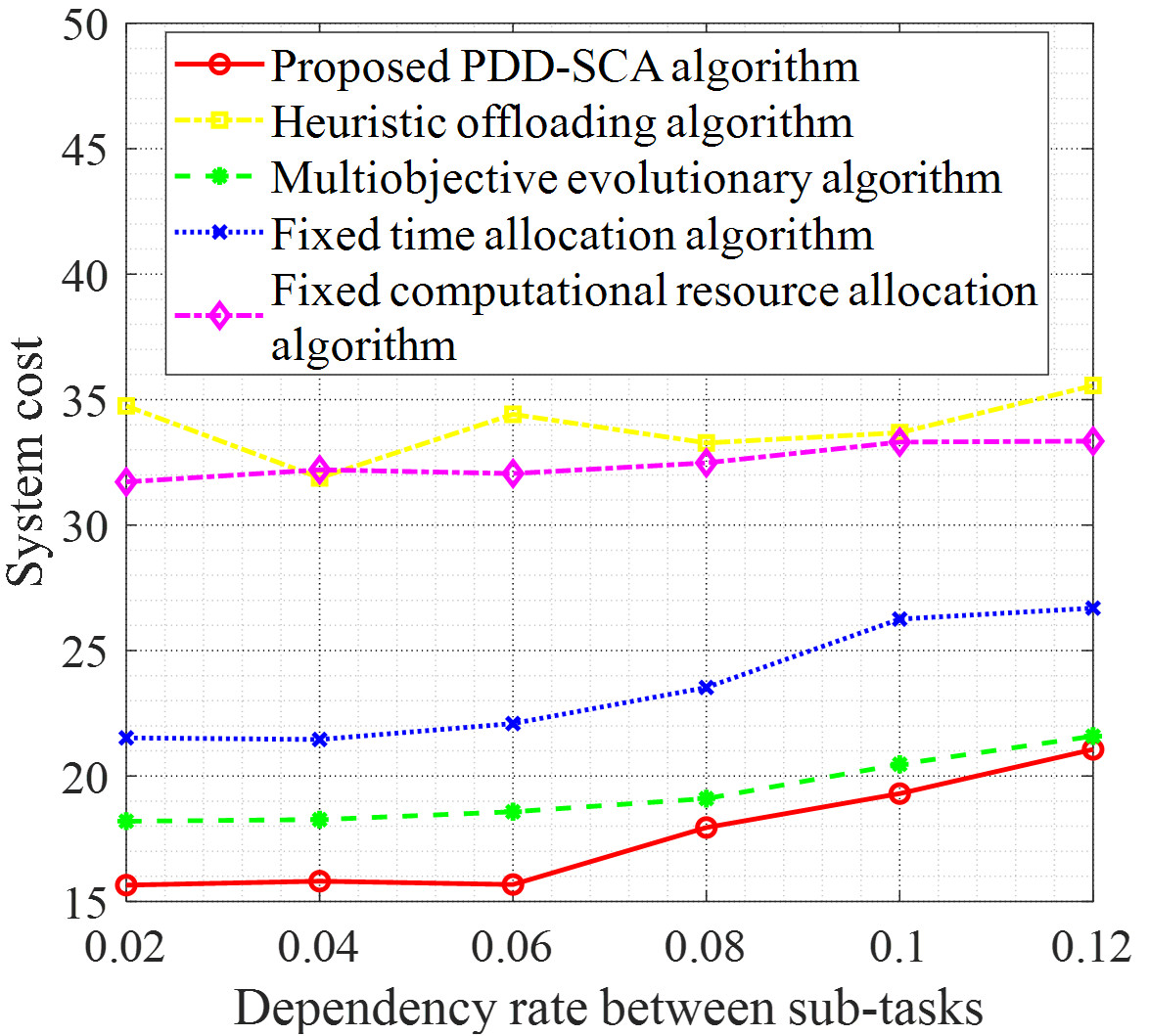}
\label{conn_sys}}%
\hfil
\subfloat[Task completion delay]{\includegraphics[width=2.28in]{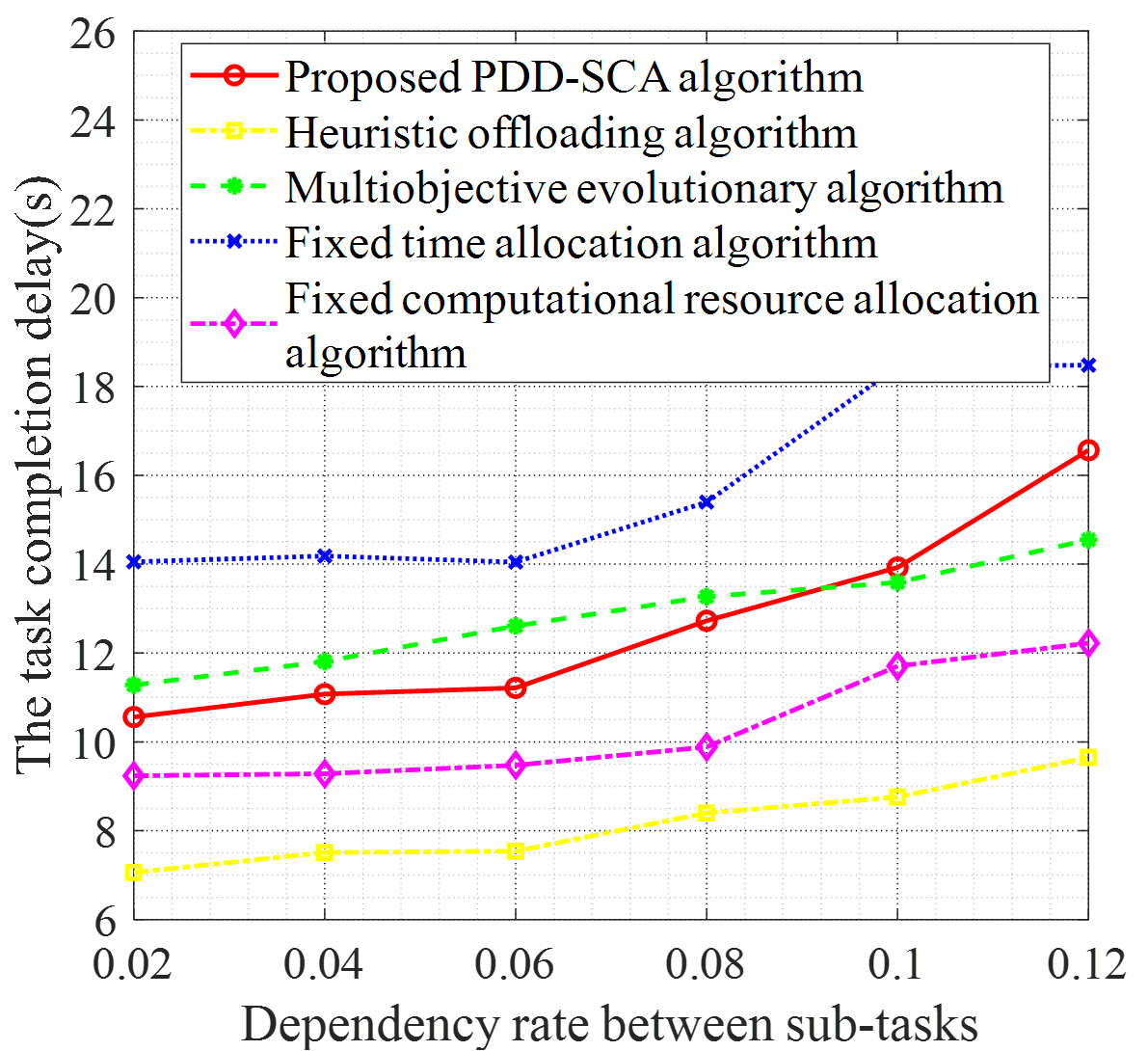}
\label{conn_delay}}%
\hfil
\subfloat[User energy consumption]{\includegraphics[width=2.32in]{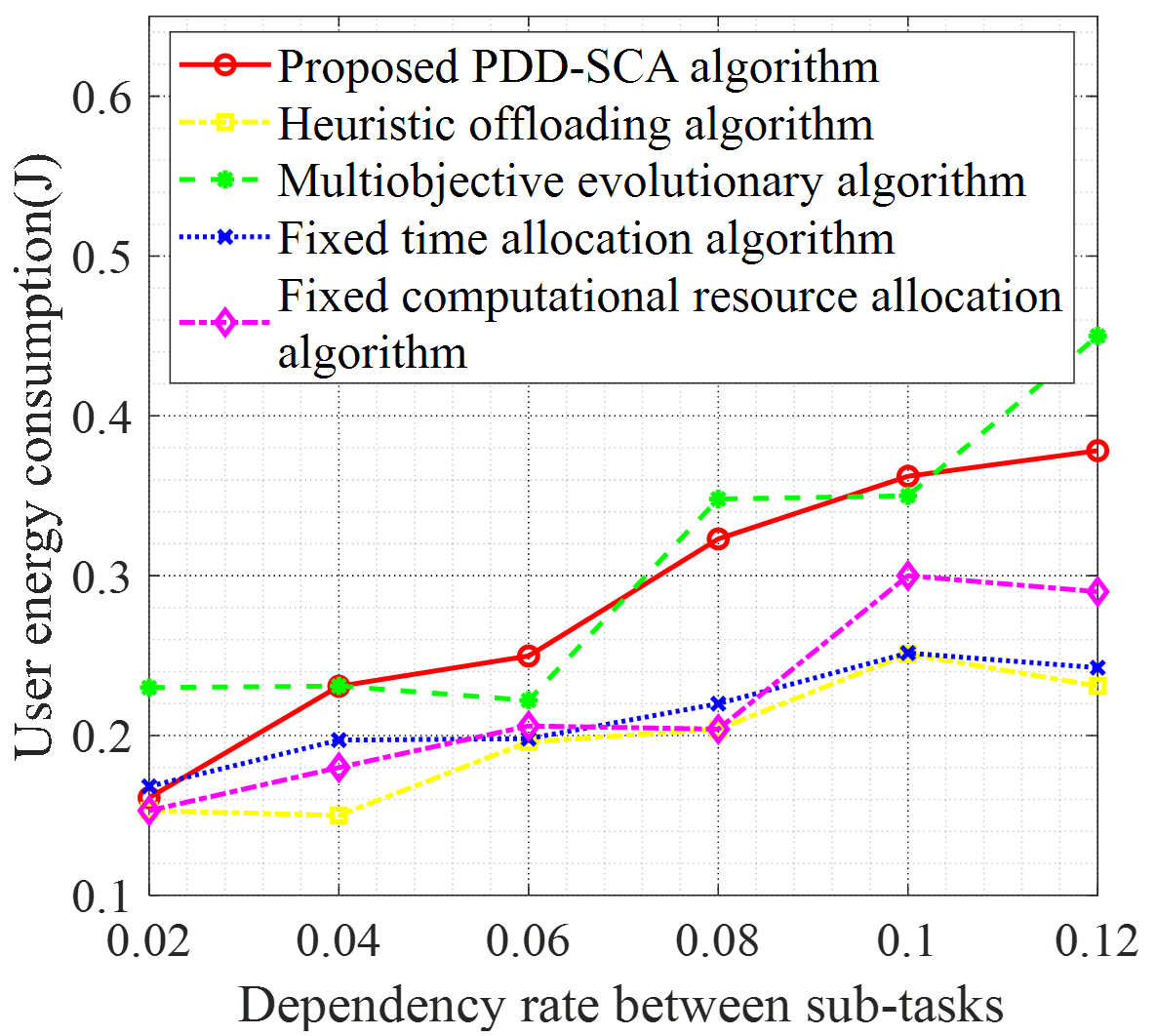}
\label{conn_eu}}%
\caption{System performance with varying inter-sub-task dependency rates.}
\label{fig7}
\end{figure*}

Fig. \ref{fig5} shows the system performance trends under different computational workloads. Specifically, we present the variation trend of the system cost, which reflects the overall system performance of the algorithm, as well as the trends of its constituent components, including task completion delay, user energy consumption, UAV computational and communication energy consumption, and UAV flight energy consumption. This decomposition not only facilitates understanding of how each component contributes to the overall system performance, but also enables more intuitive comparisons under different algorithms and workload conditions. In Fig. \ref{fig5}, as the computational workload increases, both computation delay and energy consumption also rise. Consequently, this results in a rise in system cost, task completion delay, UAV computational and communication energy consumption, as well as UAV flight energy consumption. In detail, as shown in Fig. \hyperref[workload_delay]{5(b)} and Fig. \hyperref[workload_em]{5(d)}, the proposed PDD-SCA algorithm and the multiobjective evolutionary algorithm achieve the lowest system cost and UAV computational and communication energy consumption compared to other algorithms, as they can flexibly adjust offloading decisions and computational resource allocation. Among them, the PDD-SCA algorithm outperforms the multiobjective evolutionary algorithm by jointly optimizing offloading decisions, computational resource allocation, and UAV flight trajectories. For the other three algorithms, as shown in Fig. \hyperref[workload_delay]{5(b)} and Fig. \hyperref[workload_em]{5(d)}, the heuristic offloading algorithm assigns all sub-tasks of each TD to the nearest UAV and fully allocates computational resource to them, resulting in the lowest task completion delay but the highest UAV computational energy consumption. Compared to the fixed time allocation algorithm, the fixed computational resource allocation algorithm assigns more computational resource to sub-tasks, leading to lower task completion delay but higher computational energy consumption. Additionally, as shown in Fig. \hyperref[workload_eu]{5(c)}, all five algorithms exhibit similar performance in terms of user energy consumption. This is because UAVs have sufficient computational resource, and the algorithms tend to offload more sub-tasks to UAVs to optimize system cost. Consequently, total user energy consumption is primarily determined by the communication energy required to transmit input data from TDs to UAVs in the first time slot. Since the amount of data transmitted by TDs in the initial time slot remains relatively constant, the variation in user energy consumption among the considered algorithms is minimal. Furthermore, as shown in Fig. \hyperref[workload_eprop]{5(e)}, the trend of UAV flight energy consumption closely follows that of task completion delay.\par

Fig. \ref{fig6} illustrates the variations in system cost, task completion delay, and user energy consumption under varying inter-sub-task communication data volumes. In UAV operations, computational energy consumption is generally much higher than communication energy consumption. Therefore, changes in inter-sub-task communication data volume have minimal impact on UAV computation and communication energy consumption. Additionally, UAV flight energy consumption is primarily influenced by task completion delay. As a result, experimental results for UAV computational and communication energy consumption, and flight energy consumption are not included here. Furthermore, to better highlight the impact of communication data variations on system cost, we adopt more stringent communication environment parameters. Specifically, the simulation is conducted in a 400 m × 400 m rectangular area, where TDs are randomly distributed within three circular regions, each with a radius of 50 m. UAVs are deployed at the midpoint of the centerline between adjacent TD regions, maintaining a fixed flight altitude of 120 m. Additionally, as the communication data volume between sub-tasks increases, the input data volume for each task also changes accordingly. As shown in Fig. \ref{fig6}, the increase in data volume between sub-tasks leads to higher communication latency and communication energy consumption, which consequently results in increased system cost, task completion delay, and user energy consumption. Among the evaluated approaches, the heuristic offloading algorithm assigns all sub-tasks of each TD to the nearest UAV, thereby eliminating the need for inter-sub-task communication in time slots beyond the first one (when the TD transmits data to the UAV). As a result, its task completion delay increases at a slower rate compared to other algorithms as inter-sub-task data volume grows, leading to relatively minor fluctuations in system cost. Additionally, due to the random distribution of TDs, computational loads among UAVs vary, causing fluctuations in task completion delay in the heuristic offloading algorithm.\par

Fig. \ref{fig7} illustrates the variations in system cost, task completion delay, and user energy consumption under varying inter-sub-task dependency rates. The simulation environment remains consistent with that in Fig. \ref{fig6}. As shown in Fig. \ref{fig7}, similar to the impact of increasing inter-sub-task communication data volume, a higher inter-sub-task dependency rate leads to increased communication latency between devices and the corresponding rise in communication energy consumption. Consequently, this results in higher system cost, prolonged task completion delay, and greater user energy consumption.\par

\begin{figure}
    \centering
    \includegraphics[width=0.9\linewidth]{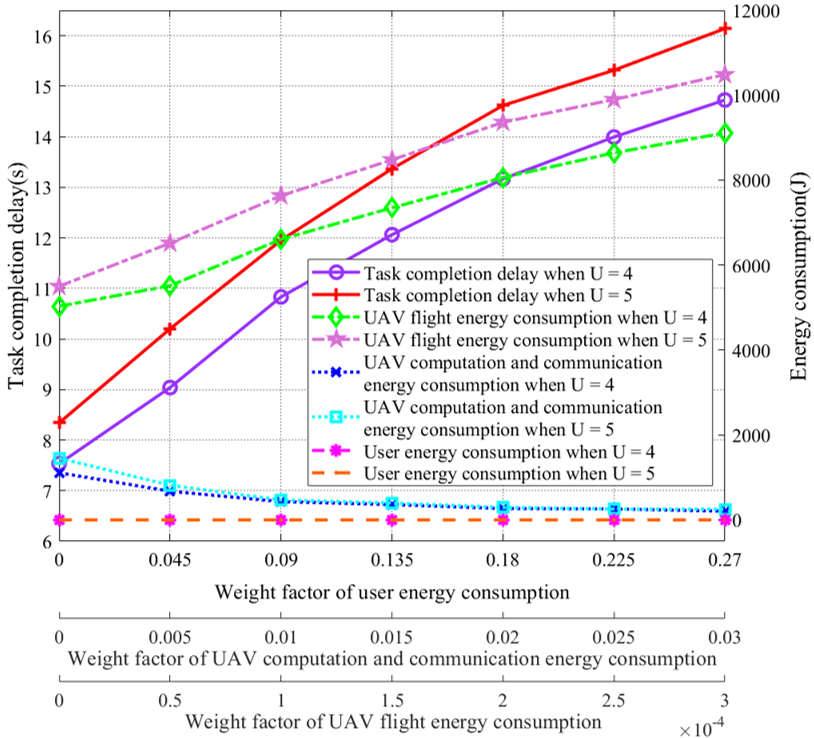}
    \caption{Task completion delay and system energy consumption with varying weight factors.}
    \label{fig10}
\end{figure}

To evaluate the trade-off between task completion delay and system energy consumption in the PDD-SCA algorithm, we analyze how varying weight parameters impact both metrics, as shown in Fig. \ref{fig10}. This experiment highlights the balance achieved between delay and energy consumption under different parameter settings. Fig. \ref{fig10} depicts the effects of different weight parameters on delay and system energy consumption for scenarios with 4 and 5 TDs. Notably, to simplify the analysis of the results, we fix the weight factor for task completion delay (i.e., $w^{\rm tim} = 1$) throughout the experiments. Although $w^{\rm tim}$ remains constant, adjusting the energy-related weights effectively alters the relative importance of delay in the system cost. As the energy weights increase, delay becomes less dominant in the objective function, and vice versa. This design facilitates clearer observation of the trade-off trends between delay and energy consumption, while avoiding unnecessary complexity in experimental interpretation. In the Fig. \ref{fig10}, it can be observed that as the weight assigned to energy consumption increases, UAV computation and communication energy consumption decrease, while the task completion delay rises. Furthermore, an increase in the weight parameter for UAV flight energy consumption still results in higher UAV flight energy consumption due to extended task completion delay. In comparison, user energy consumption remains relatively stable, as it is mainly composed of communication energy consumption in the first time slot.\par

\begin{figure}
    \centering
    \includegraphics[width=0.9\linewidth]{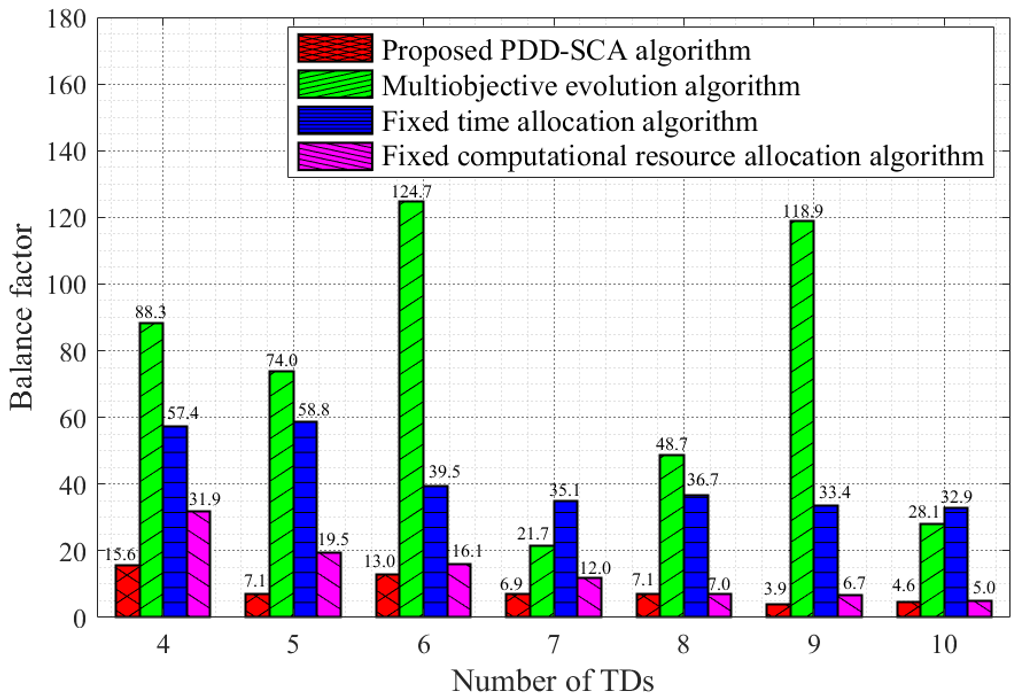}
    \caption{Balance factor with varying number of TDs.}
    \label{fig8}
\end{figure}

To verify the workload balancing performance of the PDD-SCA algorithm and benchmark algorithms, it is essential to design a balancing factor that incorporates sub-task offloading and computational resource allocation. As indicated in Eq. \eqref{eq17}, the computational energy consumption of UAVs includes both sub-task offloading and computational resource allocation. Furthermore, maintaining balanced energy consumption among UAVs is critical in practical scenarios. Thus, we employ the standard deviation of UAV computational energy consumption as the balancing factor, defined as ${(\sum\limits_{m = 1}^M {{({E_m^{\rm comp}} - \bar E)}^2}} /M) ^{1/2}$, where $\bar E = \sum\limits_{m = 1}^M {{E_m^{\rm comp}}} /M$. Due to the significantly inferior workload balancing capability of the heuristic offloading algorithm compared to the other four algorithms, it is not included in this paper's comparison. In Fig. \ref{fig8}, we graph the balance factor under varying TD numbers. As depicted in the figure, the multiobjective evolutionary algorithm has the poorest computational load balancing ability due to the more random sub-task offloading and computational resource allocation compared to other algorithms. In contrast, the proposed PDD-SCA algorithm achieves the lowest balancing coefficient through the optimization of joint sub-task offloading and computational resource allocation. Additionally, the fixed computational resource allocation algorithm, which evenly distributes sub-tasks to the UAVs and each UAV's CPU frequency to the associated sub-tasks, obtains a lower balance factor than the fixed time allocation algorithm.\par

\section{Conclusions}

This paper investigated a multi-UAV-assisted collaborative MEC framework, in which the tasks of each TD can be decomposed into sub-tasks that have serial dependencies or parallel relationships with each other. A two-timescale frame structure is designed to decouple the sub-task dependencies for dependency-aware task offloading. To minimize the system cost (i.e., the weighted sum of task completion delay and system energy consumption), a joint optimization problem of sub-task offloading, computational resource allocation, and UAV trajectories was formulated. To solve this challenging MINLP problem, the PDD-SCA algorithm was proposed. Specifically, the original MINLP problem was transformed into a more tractable form based on the PDD theory. Then, the resulting problem was decomposed into three nested subproblems, where the non-convex components were reformulated by the SCA method and the suboptimal solution was obtained by iteratively solving these subproblems. Numerical results demonstrated the superiority of the proposed algorithm in terms of system cost and workload balancing. Furthermore, a favorable trade-off between delay and energy consumption can be achieved. \par

In future research, extending the scenarios discussed in this paper to large-scale networks with a substantial increase in device numbers should be considered, where it is crucial to study efficient distributed real-time algorithms. Particularly for urgent or time-sensitive applications, distributed real-time task offloading remains a critical area requiring further investigation.


{
\bibliographystyle{IEEEtran}
\bibliography{IEEEabrv,bibe}
}

\end{document}